%% file: MY71.tex
\documentstyle[psfig]{mn}

\newcommand{\relmodv}{$\langle|V|\rangle/S$}
\newcommand{\relv}{$\langle V\rangle/S$}

\title
{Circular polarization in pulsar integrated profiles}

\author[J.~L. Han et al.]
       {J.~L. Han$^{1}$, %\thanks{Email:hjl@class1.bao.ac.cn}
	R. N. Manchester$^2$, %\thanks{Email:rmanches@atnf.csiro.au}
	R.~X. Xu$^{3}$,
	and
	G.~J. Qiao$^{4,3,5}$ %\thanks{Email:gjn@pku.edu.cn}
\\
$^1$Beijing Astronomical Observatory, Chinese Academy of Sciences (CAS),
	Beijing 100012, China \\
$^2$Australia Telescope National Facility, CSIRO, PO Box 76, Epping NSW
       2121, Australia\\
$^3$Department of Geophysics, Peking University (PKU), Beijing 100871, China.\\
$^4$CCAT (World Laboratory), PO Box 8730 Beijing 100080, China\\
$^5$Beijing Astronomy Center, CAS-PKU, Beijing 100871, China
 }

\begin{document}

\maketitle

%\label{firstpage}

\begin{abstract}
We present a systematic study of the circular polarization in pulsar
integrated profiles, based on published polarization data. For core
components, we find no significant correlation between the sense-change of
circular polarization and the sense of linear position angle variation.
Circular polarization is not restricted to core components and, in some
cases, reversals of circular polarization sense are observed across
the conal emission. In conal double profiles, the sense of circular
polarization is found to be correlated with the sense of position-angle
variation. Pulsars with a high degree of linear polarization often have one
hand of circular polarization across the whole profile. For most pulsars,
the sign of circular polarization is the same at 50-cm and 20-cm wavelength,
and the degree of polarization is similar, albeit with a wide scatter. 
However, at least two cases of frequency-dependent sign reversals are
known. This diverse behaviour may require more than one mechanism to
generate circular polarization. 
\end{abstract}


\section{Introduction}

From the earliest observations (e.g. Clark \& Smith 1969), it has been clear
that individual pulses from pulsars have high linear and circular
polarization, often with a sense change of circular polarization through the
pulse. The polarization distribution diagrams of Manchester, Taylor \&
Huguenin (1975), Backer \& Rankin (1980) and Stinebring et al. (1984a,b)
show clearly the variable nature of the circular polarization. Very high
degrees of circular polarization are occasionally observed in individual
pulses, even up to 100 per cent (Cognard et al. 1996) and mean values are
typically 20 -- 30 per cent.

Integrated or average pulse profiles generally have a much smaller degree of
circular polarization (e.g. Lyne, Smith \& Graham 1971; Manchester 1971),
showing that the sign of the circular Stokes parameter $V$ (here defined to
be $I_{LH} - I_{RH}$) is variable at a given pulse phase.  Significant net
$V$ is observed in the mean pulse profiles of most pulsars. In many cases,
it is concentrated in the central part of the profile and shows a reversal
of sense near the centre, but in other pulsars the same sense is retained
throughout. It is common to identify a central peak or region of a profile
as `core' emission, and the outer parts as `conal' emission (Rankin 1983;
Lyne \& Manchester 1988). Various properties, for example, spectral index,
often vary from the central to the outer parts of a profile. Rankin (1983)
suggested that core and conal emissions have different emission mechanisms,
with circular polarization being a property of core emission only, but Lyne
\& Manchester (1988) and Manchester (1995) argued that there is merely a
gradation of properties across the whole emission beam.  Radhakrishnan \&
Rankin (1990) identified two types of circular polarization:
`antisymmetric', in which there is a sense change near the centre of the
profile, and `symmetric', where the same hand of circular polarization is
observed across the pulse profile. In pulsars with antisymmetric $V$, they
found a correlation of the direction of circular sense change with the
direction of linear position-angle (PA) swing.
\begin{figure*}
\centering
\begin{tabular}{c}
\psfig{file=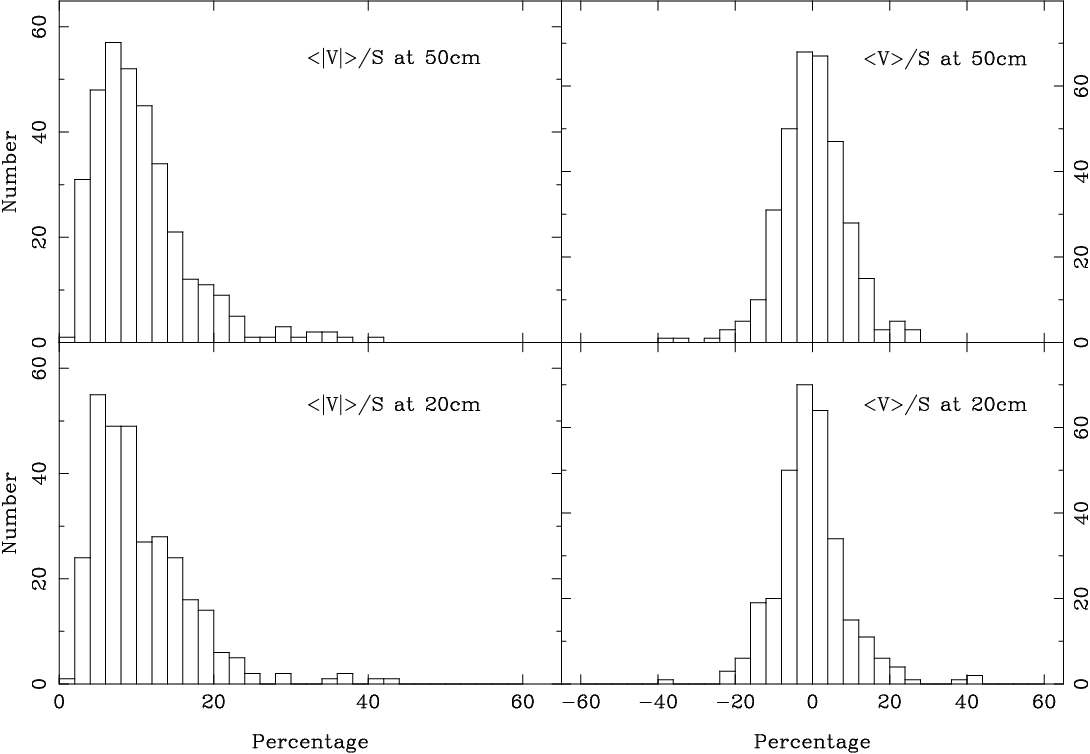,height=80mm,angle=-90}
\end{tabular}
\caption{Distribution of \relmodv~ and \relv~ at wavelengths of 50cm and 20cm.}
\label{fg:vhist}
\end{figure*}

The observed diverse circular polarization properties may relate to the
pulsar emission mechanism or to propagation effects occurring in pulsar
magnetosphere (cf. Melrose 1995). In the widely adopted magnetic
pole models (Radhakrishnan \& Cooke 1969; Komesaroff 1970; Sturrock 1971;
Ruderman \& Sutherland 1975), it seems very difficult to explain various
circular polarization behaviours. Cheng \& Ruderman (1979) suggested that
the expected asymmetry between the positively and negatively charged
components of the magnetoactive plasma in the far magnetosphere of pulsars
will convert linear polarization to circular polarization. On the other
hand, Kazbegi, Machabeli \&
Melikidze (1991, 1992) argued that the cyclotron instability,
rather than the propagation effect, is responsible for the circular
polarization of pulsars. Other authors have argued for its intrinsic origin
(Michel 1987; Gil \& Snakowski 1990a,b; Radhakrishnan \& Rankin 1990;
Gangadhara 1997). In summary, the observed circular polarization of pulsar
radiation is presently not well understood.

This paper attempts to summarize the main features of the observed circular
polarization in pulsar mean pulse profiles and to discuss these in the
context of other pulsar properties. We outline the main characteristics of
pulsar circular polarization in Section 2, discuss possible mechanisms for the
generation of circular polarization in Section 3, and present our
conclusions in Section 4.

\section{Circular polarization properties}

Information about circular polarization and PA variations from all available
published polarization profiles is summarized in Appendix A, Table A1.  Only
pulsars with significant circular polarization (i.e., signal/noise in $V$
greater than three) are included in the Table.  For about 40 per cent of the
pulsars examined, circular polarization is below this threshold.  In the
following subsections, we discuss various characteristics of the observed
circular polarization.

\subsection{Degree of circular polarization}
\begin{figure}
\centering
\begin{tabular}{c}
\psfig{file=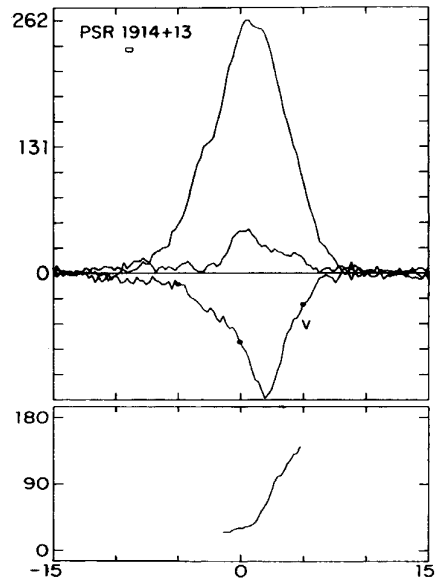,height=80mm}
\end{tabular}
\caption{Polarization profile for PSR B1914+13, a pulsar with strong
circular polarization over the whole observed profile (From Rankin,
Stinebring \& Weisberg 1989). }
\label{fg:1914}
\end{figure}

Fig.~\ref{fg:vhist} shows histograms of \relmodv~ and \relv~ in the 50-cm
band ($\sim 650$ MHz) and 20-cm band ($\sim 1400$ MHz), where $S=\langle
I\rangle$ is the mean flux density, for the pulsars in Table A1. Levels of
circular polarization are very similar at the two wavelengths; the median
value of \relmodv~ is 9 per cent at 50 cm and 8 per cent at 20 cm. Only a few
pulsars have mean circular polarization exceeding 20 per cent. The apparent
deficit of pulsars with \relmodv $\la 5$ per cent is not real; for low
signal/noise observations, noise in the profile makes a positive
contribution to \relmodv~. Although the least significant signal/noise
ratios in $V$ were not included in the sample, many of the profiles have
relatively low signal/noise ratio and hence a biased estimate of \relmodv.

As Fig.~\ref{fg:vhist} shows, strong circular polarization has been detected
from only a small number of pulsars; an example of such a pulsar is shown in
Fig.~\ref{fg:1914}. Pulsars with \relmodv $\;>20$ per cent and polarization
error $<5$ per cent are listed in Table~\ref{tb:highv}. PSR B1702$-$19, an
interpulse pulsar, has the highest known fractional circular
polarization, up to 60 per cent (Biggs et al. 1988); note, however, that
Gould (1994) gives values of 30 -- 35 per cent. PSR B1914+13, a pulsar
with almost the same $P$ and $\dot{P}$, and hence the same estimates of
characteristic age, surface magnetic field and total energy-loss rate
$\dot{E}$ as PSR B1702$-$19, also has very strong circular polarization
(Fig.~\ref{fg:1914}).
\begin{table}
\caption{Pulsars with strong circular polarization }
\label{tb:highv}
\setlength{\tabcolsep}{1.2mm}
\begin{tabular}{lrcrrrl}
\hline
PSR     & \multicolumn{1}{c}{Freq.} & \multicolumn{1}{c}{\relmodv} & 
            \multicolumn{1}{c}{\relv} & 
            \multicolumn{1}{c}{$\langle L\rangle /S$} & Err. & Ref.   \\
            & (MHz)&  (\%) & (\%) & (\%) & (\%) &  \\
\hline
B1702$-$19  &  408 &  60  & $-60$& 35 & 2 & B88 \\
B1913$+$10  & 1400 &  38  &  38  & 35 & 2 & R89\\
B1914$+$13  & 1400 &  37  & $-37$& 18 & 2 & R89\\
B0835$-$41  &  405 &  35  &   35 & 3 & 3 & H77\\
B1221$-$63  &  950 &  33  &   26 & 30 & 5 & Mc78,v97\\
B1557$-$50  & 1612 &  23  &   21 & 11 & 4 & Ma80\\
B0942$-$13  &  408 &  23  &   19 & 22 & 1 & G94\\
J1359$-$6038&  660 &  22  &   22 & 83 & 2 & M98\\
B2327$-$20  &  648 &  22  & $-22$& 16 & 1 & Mc78\\
B1737$-$30  & 1560 &  22  & $-22$& 87 & 3 & W93 \\
B1952$+$29  & 1400 &  21  & $-19$& 18 & 1 & R89\\
B1857$-$26  & 1612 &  20  & $ -1$& 23 & 3 & Ma80\\
\hline
\end{tabular}
References:
Hamilton et al. 1977 (H77);
McCulloch et al. 1978 (Mc78); 
Manchester, Hamilton \& McCulloch 1980 (Ma80);
Biggs et al. 1988 (B88); 
Rankin, Stinebring \& Weisberg 1989 (R89);
Wu et al. 1993 (W93);
Gould 1994 (G94);
van Ommen et al. 1997 (v97);
Manchester, Han \& Qiao 1998 (M98).
\end{table}

As Table~\ref{tb:highv} shows, all but one of the pulsars with strong
circular polarization is symmetric, i.e., with no reversals of $V$ across
the pulse profile. These pulsars are discussed further in Section 2.4 below.
Although most of these pulsars also have relatively high levels of
fractional linear polarization, a few do not. 

\subsection{Antisymmetric circular polarization and core emission}
\begin{table}
\caption{Circular sign reversals and sign of PA variation.}
\label{tb:core_rev}
\begin{tabular}{llll}
\hline
\multicolumn{2}{c}{LH/RH = $+/-$}&\multicolumn{2}{c}{RH/LH =
 $-/+$}\\
\multicolumn{1}{c}{ PA: dec} & \multicolumn{1}{c}{ PA: inc} & 
\multicolumn{1}{c}{ PA: dec} & \multicolumn{1}{c}{ PA: inc} \\
\hline                      
B1237+25     & B0823+26m    & B0329+54       & B0450$-$18  \\
B1508+55     & J1001$-$5507?& J0437$-$4715   & B1700$-$32  \\
B1737+13     & J1604$-$4909 & B0942$-$13     & J2144$-$3933\\
J1801$-$0357?& B2002+31     & B1323$-$58     &             \\
B1821+05?    & B2003$-$08   & B1451$-$68     &             \\
B1857$-$26   & B2113+14     & J1527$-$5552   &             \\
B1859+03     &              & B1534+12       &             \\
B2045$-$16   &              & J1852$-$2610   &             \\
B2111+46     &              & B1907+02       &             \\
\hline
\end{tabular}
\\
For PSRs B1821+05, J1801$-$0357 and J1001$-$5507, the PA variation is not very clear. 
\end{table}
As mentioned in Section 1, circular polarization is often stronger in
the central or core regions of a profile, and often shows a reversal in sign
near the profile centre. From a sample of 25 pulsars, Radhakrishnan \&
Rankin (1990) found a correlation between the sense of the sign reversal and
the sense of PA swing, with transitions of circular polarization from LH to
RH ($+/-$) being associated with decreasing PA (clockwise rotation on the
sky) across the profile, and vice versa. The proposed correlation was
questioned by Gould (1992, 1994) who found some contrary examples.

\begin{figure}  
\centering
\begin{tabular}{c}
\null\hspace{5mm}
\psfig{file=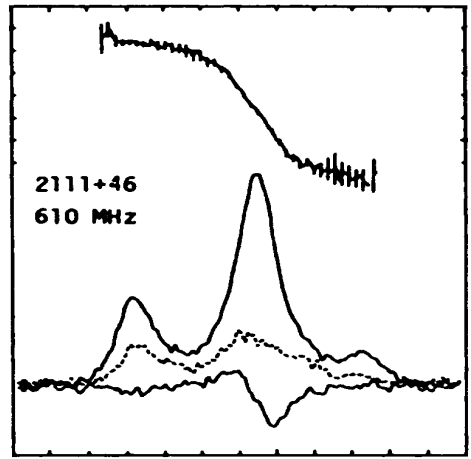,height=65mm,width=65mm} \\
\psfig{file=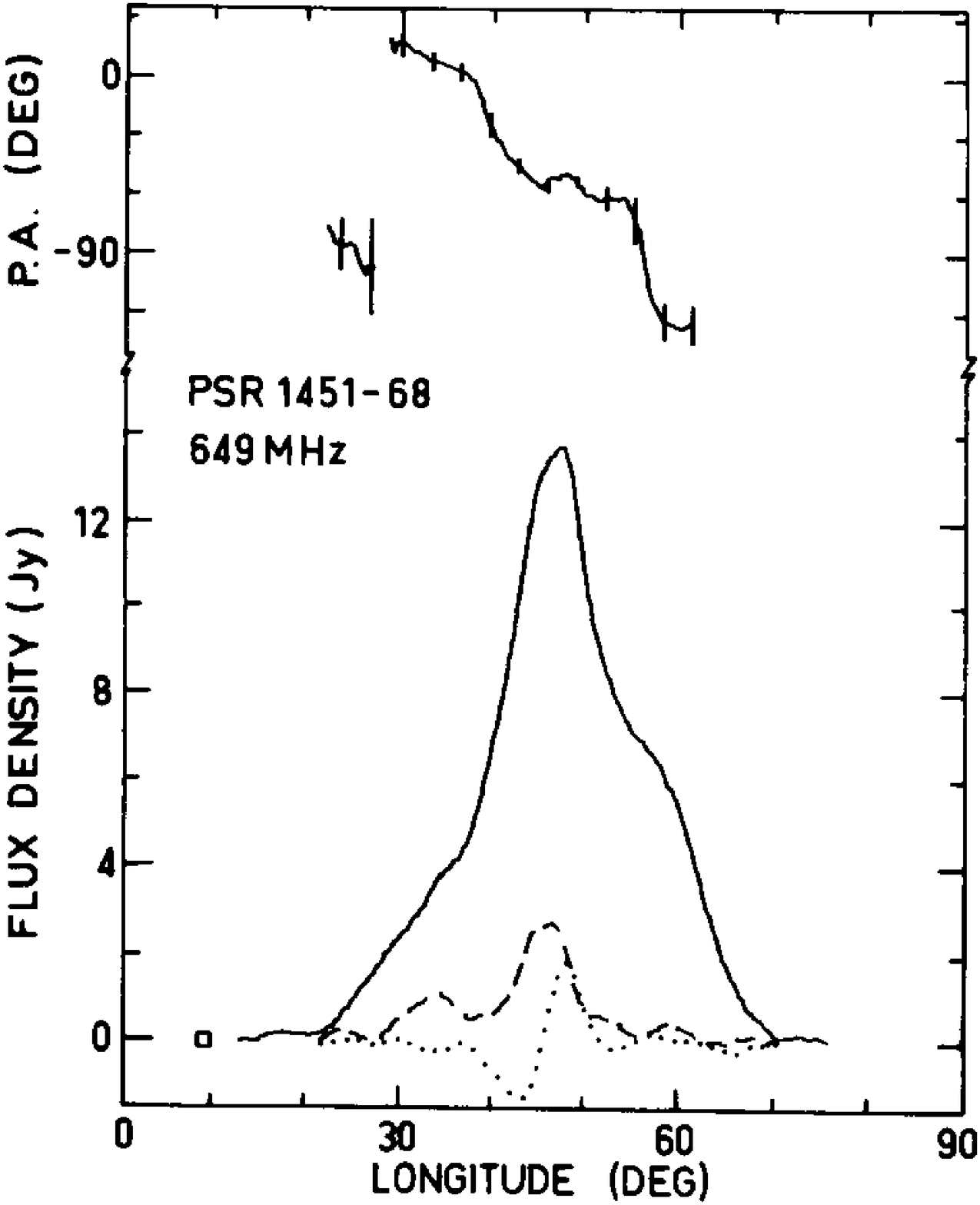,height=75mm,width=75mm,angle=0.5}
\end{tabular}
\caption{Two examples of pulsars with central reversals in the sign of
$V$. (From Lyne \& Manchester 1988 and McCulloch et al. 1978)}
\label{fg:core_rev}
\end{figure}

Table~\ref{tb:core_rev} lists pulsars with a reversal in the sign of $V$ in
the central or core region of the profile, and two good examples are shown in
Fig.~\ref{fg:core_rev}.  In Table~\ref{tb:core_rev}, pulsars are divided
into different columns according to the sense of the $V$ sign change and the
direction of PA swing.  Further information about the observations can be
found in Table A1.  Among 27 pulsars listed, 12 pulsars (in the first and
fourth columns) agree with the correlation proposed by Radhakrishnan \&
Rankin (1990) and 15 pulsars (in the second and third columns) disagree. We
therefore conclude that there is no correlation between the sense of the
sign change of circular polarization and the sense of variation of linear
polarization angle across the profile. Consequently, theoretical discussions
based on this correlation (e.g. Radhakrishnan \& Rankin 1990; Kazbegi et
al. 1991) are not well founded.

It is worth noting that central or `core' reversals in circular polarization
sense usually occur very close to the mid-point of the pulse profile and do
not appear to be related to any particular peak in the total intensity
profile. For example, PSR B0149$-$16 at 50 cm \cite{qmlg95} has a central
reversal in V between two overlapping pulse components; PSRs B1855+09 (main
pulse) and B1933+16 \cite{srs+86,rsw89} are similar. In other pulsars,
e.g. PSRs B1451$-$68 and B2003$-$08, the central peak is offset in longitude
from the profile mid-point. Some conal double pulsars, such as PSR
B2048$-$72, have a clear central reversal but no core component at
all. These observations suggest that the `core' sense reversal is associated
with the central region of the beam, rather than with any particular
component.

\subsection{Circular polarization in cone-dominated pulsars}
\begin{figure}
\centering
\begin{tabular}{c}
\psfig{file=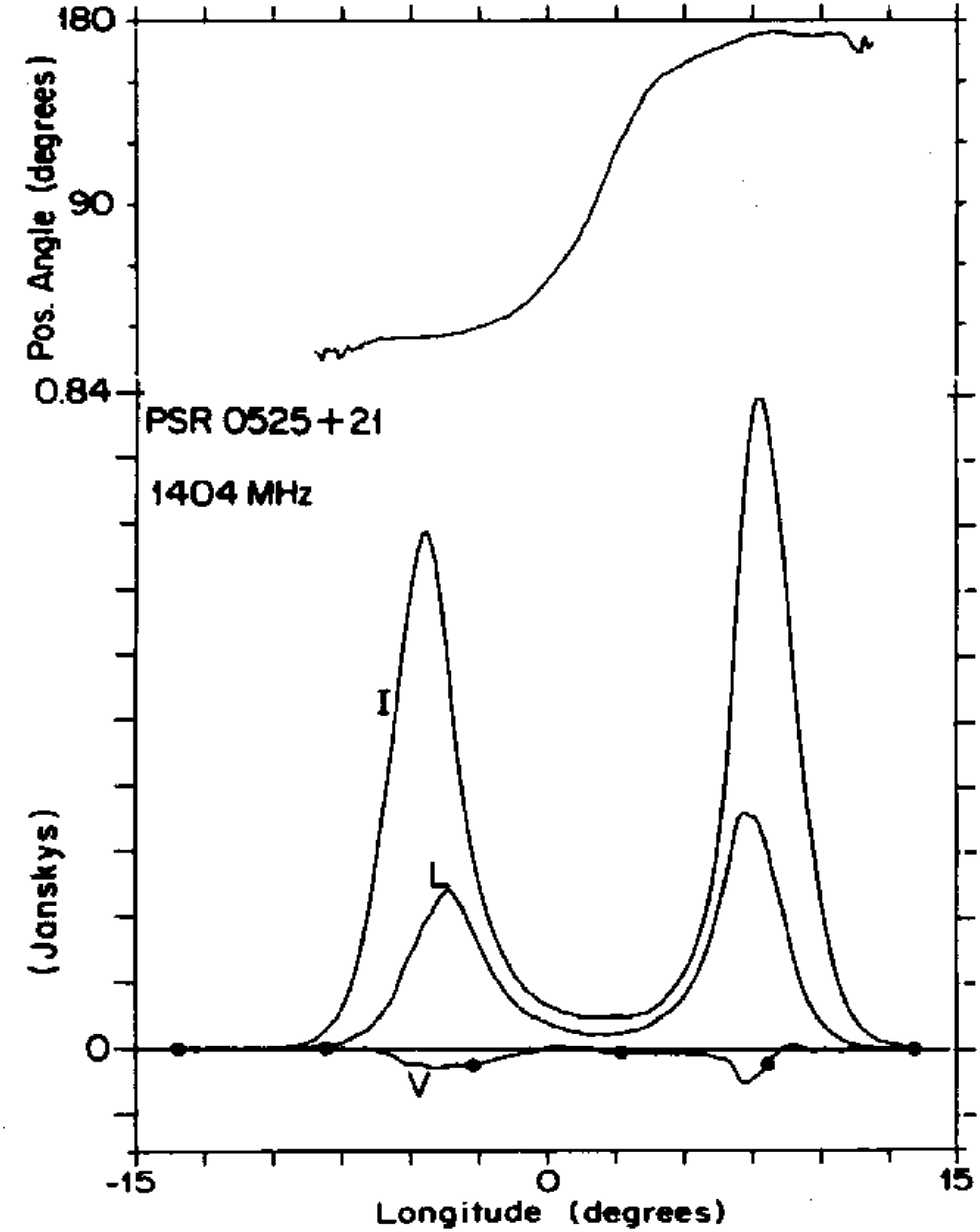,height=65mm,width=57mm} \\
\psfig{file=2346.ps,height=65mm,width=55mm}
\end{tabular}
\caption{Two examples of conal-double pulsars with significant circular
polarization. (From Stinebring et al. 1984a and Manchester, Han \& Qiao 1998)}
\label{fg:cone_double}
\end{figure}

It is often stated that the conal components of pulsar profiles have weak or
no circular polarization (e.g. Radhakrishnan \& Rankin 1990; Gil et
al. 1995). However, there are many examples of significant circular
polarization in profiles, or parts of profiles, which are believed to be
conal emission. Fig.~\ref{fg:cone_double} shows two `conal-double' pulsars
which show significant circular polarization. 

In most cases, the circular polarization is symmetric, i.e., it has the same
sign over the profile. In Table~\ref{tb:cone_double}, we show the
relationship between the sign of PA variation and sense of circular
polarization in conal-double pulsars. A striking conclusion is that there is a
strong correlation between these two properties, with right-hand (negative)
circular polarization accompanying increasing PA and vice versa. No good
examples contrary to this trend have been found. Therefore, unlike the
correlation between the sense of sign change in core emission and PA swing,
this correlation appears to be significant.

\begin{table}
\caption{Conal-double pulsars with significant circular polarization.}
\label{tb:cone_double}
\begin{tabular}{llccl}
\hline
PSR    & PA  &  \multicolumn{2}{c}{Sign of $V$} & Ref.\\
            &     & Comp 1 & Comp 2 & \\
\hline 
B0148$-$06  & inc & $-$ & $-$ & LM88 \\
B0525+21    & inc & $-$ & $-$ & S84a,R83\\
B0751+32    & inc & $-$ & $-$ & R89 \\
B0818$-$13  & inc & $-$ & $-$ & v97,Q95,B87\\
B0834+06    & inc & $-$ & $-$ & Mc78,S84a \\
            &     &$+/-$& $-$ & S84b\\
B1133+16    & inc & $-$ & $-$ & Mc78,S84a\\
B1913+16    & inc &$-/+$& $-$ & C90 \\
B2020+28    & inc &$+/-$& $-$ & C78,S84a\\
B2044+15    & inc & $-$ & $-$ & G94 \\
B2048$-$72  & inc &$\pm$ &$\mp$ & Q95,M98 \\
B0301+19    & dec & $+$ & $+$ & R83,R89\\
J0631+1036  & dec & $+$ & $+$ & Z96 \\
B1039$-$19  & dec & $-/+$ & $+$ & LM88,G94 \\
J1123$-$4844& dec & $+$ & $+$ & M98 \\  
B1259$-$63  & dec & $+$ &$\cdot$& MJ95 \\
J1527$-$3931& dec & $+$ & $+$ & M98  \\
B1727$-$47  & dec & $+$ &$\cdot$& H77,Mc78,v97\\  
J1751$-$4657& dec &$-/+$& $+$ & M98  \\
B2321$-$61    & dec & $+$ &$\cdot$& Q95\\
J2346$-$0609& dec &$\cdot$&$+$& M98 \\   
\hline
\end{tabular}
References:
Hamilton et al. 1977 (H77);
Cordes, Rankin \& Backer 1978 (C78);
McCulloch et al. 1978 (Mc78);
Rankin 1983 (R83);
Stinebring et al. 1984a (S84a); Stinebring et al. 1984b (S84b);
Biggs et al. 1987 (B87);
Lyne \& Manchester 1988 (LM88);
Rankin, Stinebring \& Weisberg 1989 (R89);
Cordes, Wasserman \& Blaskiewicz 1990 (C90);
Gould 1994 (G94);
Manchester \& Johnston 1995 (MJ95);
Qiao et al. 1995 (Q95);
Zepka, Cordes \& Wasserman 1996 (Z96);
van Ommen et al. 1997 (v97);
Manchester, Han \& Qiao 1998 (M98).
\end{table}

In some cases, a change in the sign of $V$ is observed, generally associated
with the first component. Exceptions are PSR B2048$-$72, where it is
centrally located, and PSR B0329+54 in its `abnormal' mode, where there is a
sense change in the trailing conal component (Xilouris et al. 1995). Note
also that, for PSR B0834+06, there is a sign change under the first
component at 800 MHz, but not at 1404 MHz (Stinebring et al. 1984a,b).

In summary, these observations show that circular polarization and sense
changes are not associated with any particular `core' component. The
implication is that there is no fundamental difference between core and
conal emission.
\begin{figure} 
\centering
\begin{tabular}{c}
\psfig{file=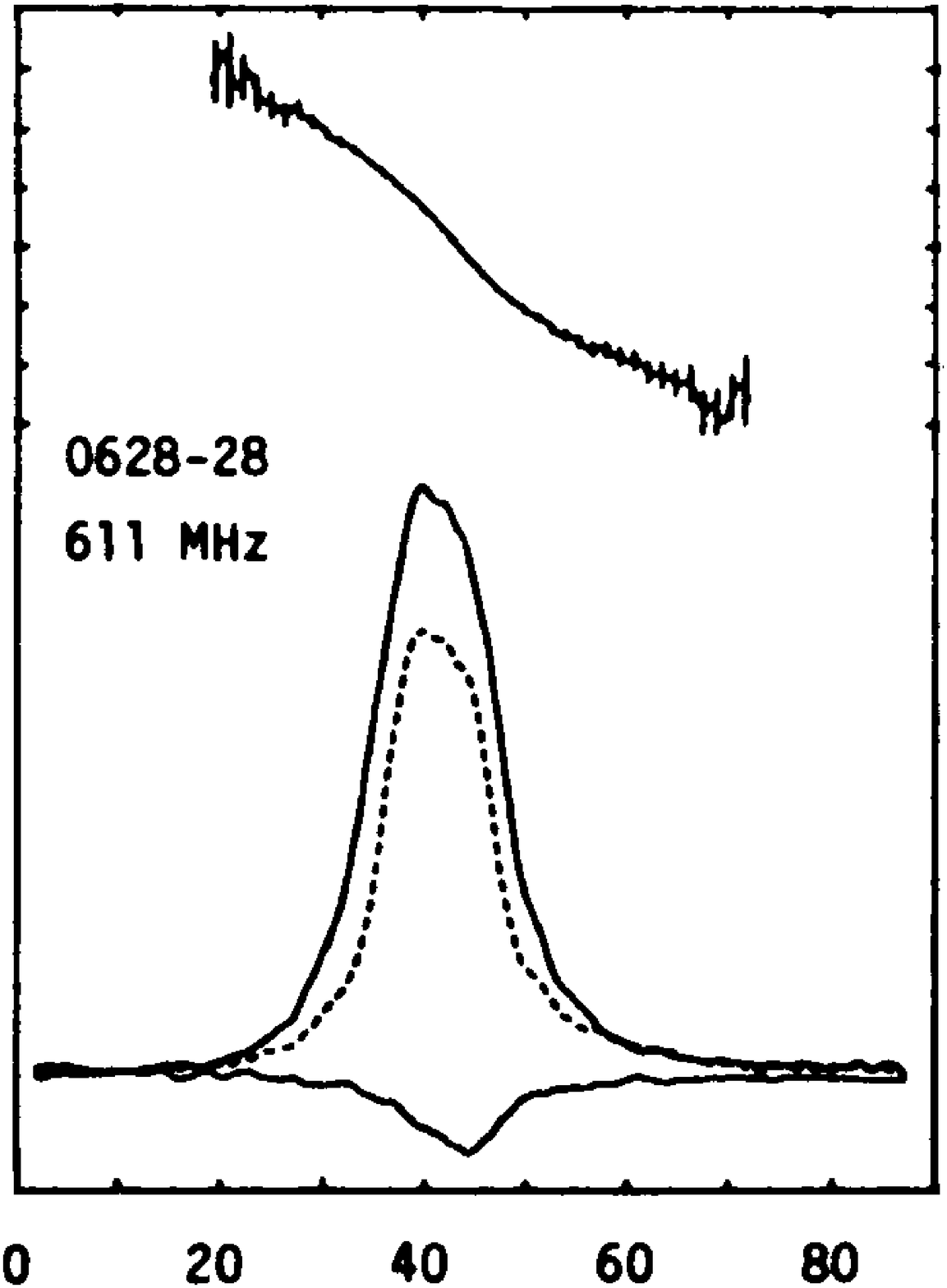,height=70mm} 
\end{tabular}
\caption{PSR B0628$-$28, an example of a pulsar with `symmetric' circular
polarization and high linear polarization. (From Lyne \& Manchester 1988)}
\label{fg:0628}
\end{figure}

\subsection{Symmetric circular polarization}
For a number of pulsars, we see one hand of circular
polarization over the whole profile, the so-called `symmetric'
type by Radhakrishnan \& Rankin (1990). 
Table~\ref{tb:symm} lists pulsars of this type; see Table
A1 for more observation details and references. Most of these pulsars
have one component or several unresolved components, and a large majority
have strong linear polarization. A good example, PSR B0628$-$28, is
shown in Fig.~\ref{fg:0628}. 

\begin{table}
\caption{Pulsars with symmetric circular polarization}
\label{tb:symm}
\begin{tabular}{llll}
\hline
\multicolumn{2}{c}{LH = +} & \multicolumn{2}{c}{RH = $-$}  \\
\multicolumn{1}{c}{PA: inc} & \multicolumn{1}{c}{PA: dec}   &  
\multicolumn{1}{c}{PA: inc} & \multicolumn{1}{c}{PA: dec}  \\
\hline
B0611+22     & B1913+10?&J0134$-$2937 & B0538$-$75? \\
B0835$-$41   &          &J1603$-$5657 & B0540+23    \\
J0942$-$5657 &          &B1706$-$14   & B0628$-$28  \\
J1202$-$5820 &          &B1914+13     & B0740$-$28  \\
J1253$-$5820 &          &B2315+21     & B0833$-$45  \\
J1359$-$6038 &          &B2327$-$20   & B0950+08    \\
J1722$-$3712 &          &             & B1629$-$50? \\
             &          &             & B1702$-$19m \\
             &          &             & B1702$-$19i \\
             &          &             & B1737$-$30  \\
             &          &             & B1915+13    \\
             &	        &             & B1937$-$26  \\
\hline 
\end{tabular}
\end{table}

It is worth mentioning that receiver cross-coupling can result in spurious
circular polarization in strongly linearly polarized sources (e.g. Rankin \&
Benson 1981; Thorsett \& Stinebring 1990). However, we believe that
this is not relevant for most of the pulsars listed here, and hence that the
correlation of symmetric circular polarization with strong linear
polarization is real.

The emission from these pulsars is probably conal, either from a grazing cut
of the emission beam (as in PSR B0628$-$28), or from a leading or trailing
component. Many of these pulsars are in the young and highly polarized class
identified by Manchester (1996) as having very wide conal beams. As shown
above, conal-double pulsars, for which the emission comes from beam edge,
generally have a single sense of circular polarization over each
component. However, Table~\ref{tb:symm} shows that the correlation between
sign of $V$ and sense of PA swing observed in conal-double pulsars is not
seen in the symmetric-$V$ pulsars. 

\begin{figure} 
\centering
\begin{tabular}{c}
\psfig{file=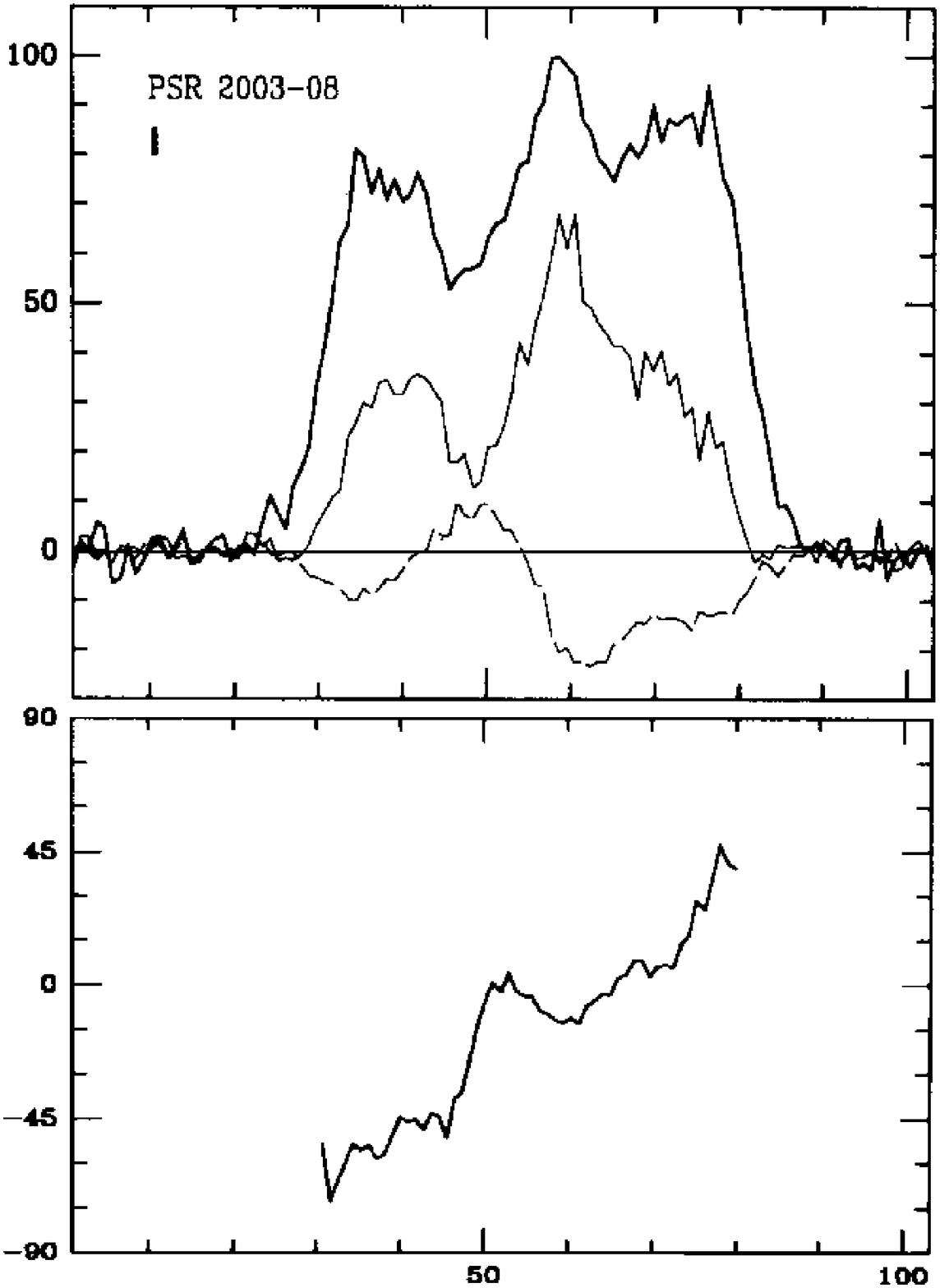,height=80mm} 
\end{tabular}
\caption{PSR B2003$-$08, a pulsar showing both symmetric and antisymmetric
circular polarization. (From Xilouris et al. 1991)}
\label{fg:2003}
\end{figure}
\begin{figure}
\centering
\begin{tabular}{c}
\psfig{file=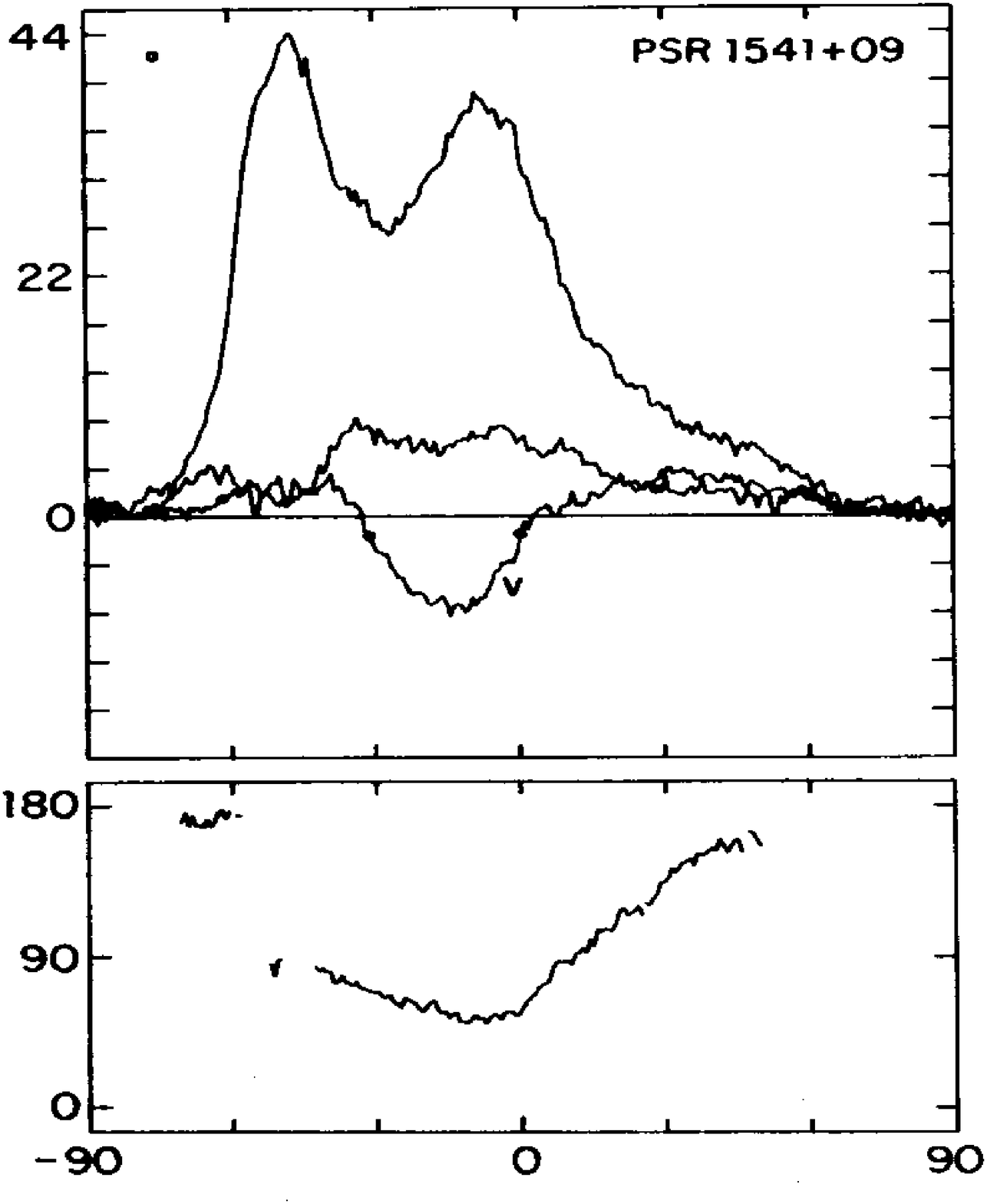,height=75mm}
\end{tabular}
\caption{PSR B1541+09, a pulsar with a complex variation of circular
polarization across the profile. (From Rankin et al. 1989)}
\label{fg:1541}
\end{figure}

\subsection{Complex variations of circular polarization with longitude}
As described above, circular polarization from central or core regions of
the profile is usually antisymmetric, whereas that from conal regions is
usually symmetric. In some pulsars, e.g. PSR B2003$-$08 (Fig.~\ref{fg:2003})
these two properties are superimposed, resulting in a more complex pattern
of $V$ sign changes across the profile.

Complex variations, possibly related to different components of the profile,
are seen in PSR B1541+09 (Fig.~\ref{fg:1541}) and PSR B1952+29 (Rankin et
al. 1989). PSR B1907+10 (Rankin et al. 1989) has relatively strong RH
circular polarization, apparently related to one component of the profile.

Without doubt, the most complex variation of circular (and linear)
polarization known is seen in PSR J0437$-$4715 (Navarro et al. 1997). As
with many other pulsars, circular polarization is strongest near the main
central peak of the profile and has a sign reversal at or near the profile
centre. However, circular polarization is detected over most of the very
wide pulse profile, with several sense reversals at other longitudes. On the
leading side of the profile, these reversals may be centred on pulse
components, but this is not the case on the trailing side. 

\begin{figure}
\centering
\begin{tabular}{c}
\psfig{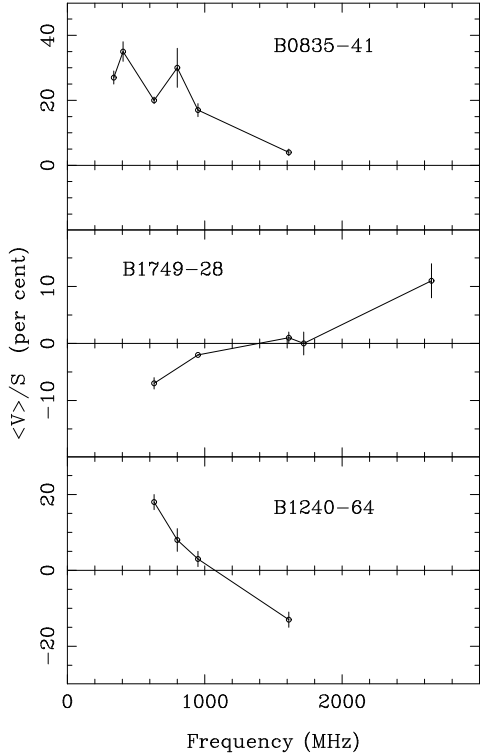}
\end{tabular}
\caption{Frequency dependence of \relv~ for three pulsars}
\label{fg:vfreq}
\end{figure}
\subsection{Frequency dependence of circular polarization}

\begin{figure*}
\centering
\begin{tabular}{c}
\psfig{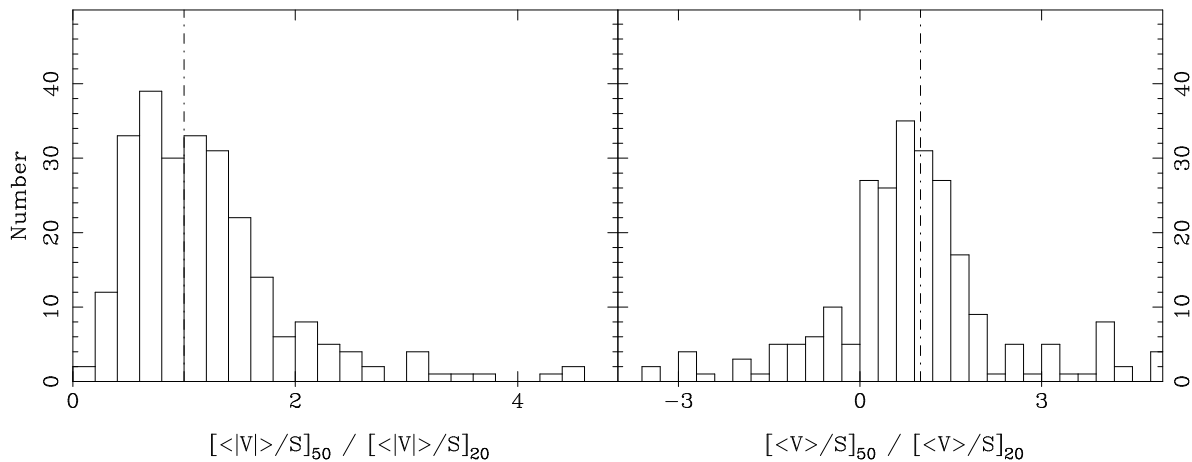}
\end{tabular}
\caption{Distribution of ratios of \protect\relmodv~ and \protect\relv~ at
wavelengths of 50 cm and 20 cm. The dot-dashed line marks a ratio of 1.0.}
\label{fg:vratio}
\end{figure*}

As discussed in Section 2.1, there is little systematic difference in the
degree of circular polarization between frequencies of 650 MHz (50 cm) and
1400 MHz (20 cm). This is further illustrated in Fig.~\ref{fg:vratio} which
shows the distributions of ratios of \relmodv~ and \relv~ at 50 and
20 cm. In both cases, the distributions are centred on 1.0 (in fact, the
median value of [\relmodv]$_{50}$/[\relmodv]$_{20}$ is exactly 1.0), showing
that there is little or no systematic variation in degree of circular
polarization with frequency.

However, these histograms do have a wide spread. A significant number of
[\relv]$_{50}$/[\relv]$_{20}$ values are negative, implying a change in sign
of $\langle V\rangle$ with frequency. Fig.~\ref{fg:vfreq} shows the
frequency dependence of \relv~ in more detail for three pulsars. For two of
these, there is clear evidence for a change in sign of $\langle V\rangle$. 
\begin{table}
\caption{Pulsars showing frequency-dependent circular polarization}
\label{tb:vfreq}
\begin{tabular}{cccl}
\hline
PSR  &low $\nu$&high $\nu$& Ref. \\
\hline
B0148$-$06 &  $-$   & none?     & LM88,W93 \\
B0149$-$16 &  $+-$  & none?     & Q95,X91\\
B0355$+$54 &  weak  & strong    & LM88,X91,X95\\
B0833$-$45 &  weak  & strong    & H77,Ma80,KD83\\
B0834$+$06 & $+/--$ & $-+-$     & S84ab,v97,Ma80\\  
B0835$-$41 & strong & weak      & H77,Mc78,Ma80\\
B0932$-$52 & $-$    & +?        & v97,M98\\
B1240$-$64 & +      & $-$       & Mc78,v97,Ma80\\
B1641$-$68 & strong & none?     & Q95,v97,W93\\
B1700$-$32 &  none? & $-/+$     & LM88,Ma80\\
B1727$-$47 & weak   & none?     & H77,Mc78,v97,Ma80\\  
B1749$-$28 & $-$    & $-/+$     & Mc78,v97,Ma80,Mr81\\
B2048$-$72 & $+/-$  & $-/+$     & Q95\\
\hline
\end{tabular}
\\
References: 
Hamilton et al. 1977 (H77);
McCulloch et al. 1978 (Mc78);
Manchester, Hamilton \& McCulloch 1980 (Ma80);
Morris et al. 1981 (Mr81);
Krishnamohan \& Downs 1983 (KD83);
Stinebring et al. 1984a (S84a); Stinebring et al. 1984b (S84b);
Lyne \& Manchester 1988 (LM88);
Xilouris et al. 1991 (X91);
Wu et al. 1993 (W93);
Qiao et al. 1995 (Q95);
Xilouris et al. 1995 (X95);
van Ommen et al. 1997 (v97);
Manchester, Han \& Qiao 1998 (M98).
\end{table}

Table~\ref{tb:vfreq} lists pulsars with apparently significant
frequency-dependent circular polarization. In some cases, such as those
shown in Fig.~\ref{fg:vfreq}, there seems no doubt about the reality of the
frequency dependence. In other cases, further observations are required to
establish these trends with more certainty. Simultaneous dual-frequency
observations would be especially interesting.

\section{Discussion} 
We have presented a summary of all published observations relating to
circular polarization of pulsar mean pulse profiles. We emphasize that we
have not considered polarization of individual pulses. While individual
pulse polarization may be closely related to the generation process,
published results are not as complete and the properties are very variable,
making analysis difficult. In general, characteristics of the mean pulse
profile are stable in time, and hence provide a framework for understanding
the processes related to generating the polarization.

Generally, in astronomy, wherever there is appreciable
asymmetry, there is likely to be polarization at some level (Tinbergen 1996,
p.27). If the asymmetry is of a scalar kind (e.g. a longitudinal component of
the magnetic field), the polarization or birefringence will be circular and
scalar in character. If the asymmetry is of a vector type (e.g. a transverse
magnetic component), the polarization or birefringence will be linear and
vector in character.

For pulsars, the radio emission is believed to be generated by highly
relativistic particles moving along magnetic field lines above the magnetic
poles. These field lines are curved, giving the asymmetry a vector component
which, from the general principle above, results in linear polarization as
in the widely accepted rotating-vector model (Radhakrishnan \& Cooke 1969).
There are strong curved magnetic fields, not only in the emission
region, but also along the path in the pulsar magnetosphere through which
the radiation propagates. The emission region is very probably asymmetric
and the emission beam is anisotropic. Hence, the observed circular
polarization can be either intrinsic to the emission process or due to
propagation effects, or perhaps both.

\subsection{Intrinsic origin?}

If the observed circular polarization is intrinsic to the emission region or
radiation process, the radiation should suffer negligible modification by
propagation effects. With one or two possible exceptions, the sense-reversal
of circular polarization seen in central or core regions of pulsar profiles
is not frequency-dependent, which suggests that the circular polarization
does not arise from a propagation effect or plasma emission process (Michel
1987; Radhakrishnan \& Rankin 1990). An origin related to asymmetry in the
beam shape seems more likely. 

If the extent of the radio emission beam is determined by the last open
magnetic field lines from the polar cap, it is axisymmetric for an aligned
rotator. However, the axisymmetry is broken for oblique rotators. For
purely geometric reasons, the beam is elongated in the longitude direction
(Biggs 1990; Roberts \& Sturrock 1972). 

Michel (1987) suggested that the currents out of a polar cap should
preferentially flow along the shortest open field lines to the light
cylinder and argued that pulsar radiation is not from a complete hollow
cone, but is concentrated in the half-cone at lower latitude.  In contrast,
in the model developed by Arons and co-workers (e.g. Arons 1983),
relativistic charged particles accelerate within the flux tubes of
``favourable curvature'' which bend towards the rotation axis. Pulsar
emission then is only possible from the upper half beam at high latitudes.

If the pulsar radio beam is randomly patchy in structure as suggested by
Lyne \& Manchester (1988) and emission is generated by relativistic
particles flowing along random flux tubes, then other asymmetries will be
present. The fact that multiple and asymmetric components are observed in
pulse profiles shows that such asymmetries exist. These may, for example,
account for the complex longitude variation of pulsar circular polarization
seen in pulsars such as PSR J0437$-$4715.

Radhakrishnan \& Rankin (1990) suggested that the elliptical shape of the
pulsar beam is responsible for the antisymmetric polarization for core
components.  According to this model, the circular polarization should be
intrinsically antisymmetric and frequency-independent. Radhakrishnan \&
Rankin (1990) argued that this mechanism could produce the correlation
between sense change and PA variation. However, as discussed in Section 2.2,
this correlation is no longer observed, indicating a different origin for
the circularly polarized emission. 

Gangadhara (1997) has suggested that positrons and electrons emit
orthogonally polarized emission as they move along curved magnetic field
lines. If the trajectories of the two species are different, coherent
superposition of the two orthogonal modes would result in antisymmetric
circular polarization. 

\subsection{Propagation effect or plasma process?}

In a relativistic electron-positron plasma, for $\theta \ll 1$,
$\omega_p^2/\omega_B^2 \ll 1$ and $\Delta \gamma / \gamma_p \ll 1$, two
circularly polarized waves can exist if
$$ \theta^2 \ll \theta_0^2 \equiv (\omega/8\omega_B)(\Delta
\gamma/\gamma_p^4) $$ 
and two linear polarization waves can exist in the opposite case, i.e.
$\theta^2\gg \theta_0^2$ (Shafranov 1967; Kazbegi et al. 1991, 1992). Here,
$\omega_B = eB/mc$, $\omega_p^2=4\pi n e^2/m$, $\theta$ is the angle between
the wave vector and the local magnetic field direction, and $\gamma_p$ is
the average Lorentz factor of the plasma.  The average difference,
$\Delta\gamma$, of Lorentz factors for $e^{\pm}$ plays a critical role in
the polarization properties.  If $\Delta\gamma$ is very small so that
$\theta^2 \gg \theta_0^2$ is always satisfied, then we have only linear
polarization modes. Conversely, when the distributions of electrons and
positrons are significantly different, circular polarization waves can exist
and propagate in the region of $\theta^2\ll \theta_0^2$.  This concurs with
the simple principle that circular polarization is related to the
longitudinal magnetic field, and allows circular polarization to propagate
near the magnetic axis where the angle between the wave vector and the
magnetic field can be always small.

Cheng \& Ruderman (1979) introduced the term ``adiabatic walking'' to
describe the slow change in the polarization properties during propagation
in the pulsar magnetosphere. This effect can result in 100 per cent
linearly polarized wave modes even when the generated waves have random
polarization. In the far magnetosphere, the conditions for adiabatic walking
no longer apply, and circularly polarized modes can be generated. Two
distinct mechanisms were discussed. One applies to a symmetric $e^{\pm}$
plasma, i.e., the positrons and electrons have the same distribution of
Lorentz factors. Adiabatic walking fails near the light cylinder if the
propagation direction and magnetic field are no longer coplanar; the
initially linearly polarized modes then become elliptical.  The second
mechanism is for an asymmetric plasma. The normal modes become elliptical when
the inequality $\omega \ll \gamma (eB/mc)$ is no longer satisfied because of
diminishing $B$. Significant circular polarization can be generated before
adiabatic walking ceases, so long as the curvature of field lines is
sufficiently small. Cheng \& Ruderman (1979) believe that this second
mechanism is likely to be more significant.

Both mechanisms predict a frequency dependence of circular polarization,
with weaker polarization at higher frequencies. As discussed in Section 2.6,
this is seen in some pulsars, but is not generally the case. Observations
over a wider frequency range would be useful.

The region of cyclotron resonance, typically located several hundred
neutron-star radii above the surface, is an interesting region, where
\[
|\omega_B|/\gamma = \omega (1-\beta \cos \theta) \simeq \omega
(\theta^2+\gamma^{-2})/2.
\]
Here $\beta=v/c$ for the particles.  Kazbegi et al. (1991)
suggested that pulsar emission is generated in this region via the cyclotron
instability. For large $\theta$, the emission is linearly polarized, but for
small $\theta$ it is circularly polarized with sense corresponding to
the charge signs. Net circular polarization can be produced by a
relativistic electron-positron plasma with $\Delta \gamma \neq 0$. This
mechanism would be expected to produce circular polarization predominantly
in the central or core regions of the profile. 

Propagation through this region may alter the polarization state of a wave.
Istomin (1992) suggested that incident linearly polarized waves become
circular as a result of generalized Faraday rotation in this
region. Radiation with a wave vector in the plane of the field line (O-mode)
becomes LH polarized and that with a wave vector normal to the plane
(E-mode) becomes RH polarized. The two circular modes have different
group velocities which could result in net circular polarization in the
observed radiation. This is a possible origin for `symmetric' circular
polarization observed in conal emission (Table~\ref{tb:symm}).

\section{Conclusions}
We have shown that, in pulsars, circular polarization is common but diverse
in nature. It is generally strongest in the central or `core' regions of a
profile, but is by no means confined to these regions. The circular
polarization often changes sense near the middle of the profile, but sign
changes are occasionally observed at other longitudes. Relatively strong
circular polarization of one hand is often observed in pulsars which also
have high linear polarization. Although examples of both increasing and
decreasing degree of circular polarization with frequency, and even changes
of sign, are observed, on average there is no systematic frequency
dependence of polarization degree. The correlation between circular
polarization sense and sense of PA swing for conal-double pulsars is very
intriguing.

It seems unlikely that these diverse behaviours can be accounted for by a
single mechanism for generation of circular polarization. Both intrinsic
emission and propagation effects seem possible.  Antisymmetric
circular emission could result from asymmetries in the emission beam or from
propagation effects in the pulsar magnetosphere. Confirmation of its frequency
independence would support an intrinsic process. On the other hand, the
strong symmetric polarization observed in conal components is most probably
generated by a propagation process. 

Further observations of the polarization of both individual and mean pulse
profiles will help to clarify many of these issues. In particular,
observations over a wide frequency range would be valuable in sorting out
the importance of propagation effects. 

\section*{Acknowledgments}

We thank Ye Jun for assistance in compiling the data presented in this
paper,
Dr. Dan Stinebring for permission to reproduce Fig.~2, 4a, \& 7
from his papers published in ApJ, and Prof. J. Lequeux for permission to
reproduce Fig.6 from A\&A.
JLH thanks the Su-Shu Huang Research Foundation for support for
travel to ATNF in 1996 January, where this work was initiated, and
acknowledges financial support from the National Natural Science Foundation
(NNSF) of China and the Research Foundation of Astronomical Committee of
CAS.  GJQ thanks the NNSF of China for support for his visits at ATNF and
acknowledges financial support from the Climbing Project -- the National Key
Project for Fundamental Research of China. JLH and GJQ acknowledge support
from the Bilateral Science and Technology Program of the Australian
Department of Industry, Science and Tourism.

\appendix
\section{The Circular Polarization Database}
Table A1 summarizes the observed circular polarization and variation of
linear position angle in mean pulse profiles for pulsars where significant
circular polarization is observed.  The first two columns give the pulsar J
name and B name (if applicable), respectively. In the third column, the sign
of $V = I_{LH} - I_{RH}$ for across the profile is indicated. Sign reversals
are indicated by $+/-$ or $-/+$ and apparently distinct polarization features are
indicated by repeated signs. The fourth column gives the sense of position
angle variation across the profile, with `inc' meaning increasing position
angle (i.e. counterclockwise on the sky) and `dec' meaning decreasing
position angle (PA). In some cases, the variation of PA is complicated by
orthogonal jumps or other reasons. Where possible, we have accounted for
this in assigning a sense to the PA variation; uncertain cases are marked
with `xx'. A few pulsars have a change in PA sense within the profile -
these are marked with, e.g. `i+d'. The following two columns are
respectively the observation frequency in MHz and the reference code,
identified at the end of the Table and in the References. 
Comments are listed in
the final column, where `$V$(f)' means that $V$ is a function of observation
frequency, and `dt' is the time resolution of the profile. For pulsars that
have been observed by several authors or at different frequencies, we list
each observation on a separate line. Where multiple observations at a
similar frequency exist, we only list the one with the best signal/noise
ratio.

\input{table_a1.tex}

\end{document}

%% file: table_a1.tex
\begin{table*}
\caption{A summary of pulsar circular polarization observations}
\setlength{\tabcolsep}{2mm}
\begin{tabular}{lllrrclrll}
\hline
\hline
\multicolumn{1}{l}{PSR J}    &
PSR B    &
\multicolumn{1}{l}{$V$}& \hspace{-7mm}\relmodv & \relv & $\sigma$ &
\multicolumn{1}{l}{PA}     &
\multicolumn{1}{r}{Freq.}  &
\multicolumn{1}{l}{Ref.}   &
\multicolumn{1}{l}{Comments}\\
 &           &    &(\%) & (\%) & (\%) &    &(MHz)&  & \\
\hline
 0034$-$0721 & 0031$-$07  &$ +      $&  &$   $&  &dec &  268&R83  &drifting subpulses.       \\
 0133$-$6957 &            &$ -+     $&10&$ -3$& 3&dec &  661&M98  &$-/+$ at center of pulse \\
 0134$-$2937 &            &$ -      $&13&$ -8$& 3&inc &  436&M98  &$-V$ over profile          \\
 0139$+$5814 & 0136+57    &$ +      $&  &$   $&  &inc &  611&LM88 &$+V$ for only comp. $L\sim 100$\%\\
             &            &$ +      $&15&$ 15$& 3&inc & 1720&X91  &2 comp seen. $+V$ near center\\
 0141$+$6009 & 0138+59    &$ +-+    $&  &$   $&  &dec &  415&LM88 &interesting profile\\
 0151$-$0635 & 0148$-$06  &$ --     $&  &$   $&  &inc &  611&LM88 &$-V$ for double cones    \\
 0152$-$1637 & 0149$-$16  &$ +-     $&12&$ -2$& 1&inc &  660&Q95  &$+V$, $-V$ for 2 comp. Large dt. $V$(f)?\\
 	     &            &$ (+)-   $& 8&$ -5$& 2&xx  &  950&v97  &no $V$ or weak? PA not clear. \\
 0206$-$4028 & 0203$-$40  &$ -      $& 8&$ -8$& 2&xx  &  660&Q95  &good detection just at one point.\\
 0211$-$8159 &		  &$ -      $&  &$   $&  &xx  &  661&M98  &$-V$ for only high peak component  \\
 0255$-$5304 & 0254$-$53  &$ -      $& 4&$ -3$& 1&xx  &  950&v97  & PA not clear, maybe dec.  \\
 0304$+$1932 & 0301+19    &$ +++    $&  &$   $&  &dec &  430&R83  &$+V$ for the cones and bridge\\
	     &            &$  +     $&  &$   $&  &dec &  430&BR80 &from individual pulses.  \\
             &	  	  &$ +++    $&10&$ 10$& 2&dec & 1400&R89  &the same as at 430 MHz.    \\
	     &            &$ +++    $&12&$ 12$& 3&dec & 1720&Mr81 &the same as 430 MHz.  \\
             &		  &$ +++    $&15&$ 15$& 4&dec & 2650&Mr81 &the same as 430 MHz.    \\
 0332$+$5434 & 0329+54    &$ (+)+/- $&  &$   $&  &dec & 1700&B82  &both modes $+/-V$ for core, V(f)  \\
	     &            &$ -/+    $&13&$ 10$& 3&dec & 1720&Mr81 &$-/+$ for core \\
             &		  &$ -/+    $&15&$  8$& 3&dec & 2650&Mr81 &$-/+$ for core \\
	     &            &$ +/-    $&10&$ -2$& 3&xx  &10550&X95  &abnormal mode: $+/-V$ for last comp\\
 0358$+$5413 & 0355+54    &$ +-+    $&  &$   $&  &dec &  415&LM88 &evolves strongly with freqency\\
	     &            &$ -+     $& 5&$ -2$& 1&dec & 1720&X91  &$-V$ for 1st comp, weak $+V$ 2nd comp\\
             &	          &$ +(-+)  $&11&$ 9 $& 3&dec & 2650&Mr81 &mode-changing.  \\
	     &            &$ --     $&14&$-14$& 2&dec &10550&X95  &$-V$ for 2 comp of both modes.\\
 0401$-$7608 & 0403$-$76  &$ -      $& 6&$ -2$& 1&dec & 1440&Q95  &$-V$ over profile.\\
 0437$-$4715 &	         &$+-+--/+++$&13&$ 8 $& 1&dec &  438&N97  & complex $V$  \\
             &	          &$---/+++ $&16&$ -1$& 1&dec &  661&N97  & complex $V$(f) \\
             &	        &$-+-+--/+-+$&11&$ -5$& 1&dec & 1512&N97  & complex $V$ \\
 0452$-$1759 & 0450$-$18  &$--/+-   $&  &$   $&  &inc &  408&LM88 &$-/+$ for core, $-V$ for cone\\
 0454$+$5543 & 0450+55    &$ +-     $&12&$-6 $& 2&dec & 1700&X91  &$V$ for second half\\
 0459$-$0210 &            &$ -      $&18&$-10$& 3&inc &  436&M98  &$-V$ for peak component    \\
 0525$+$1115 & 0523+11    &$ -+     $&  &$   $&  &xx  & 1400&R89  &$-V$, $+V$ for 2 comp.  \\
 0528$+$2200 & 0525+21    &$ -      $&  &$   $&  &inc &  430&R83  &$-V$ for one cone. Conal double\\
             & 	          &$ --     $& 3&$ -3$& 0&inc & 1404&S84a &$-V$ for both cones.  Best profile!\\
             &		  &$ -      $&  &$   $&  &inc & 1420&LM88 &$-V$ for 1st cone\\
 0536$-$7543 & 0538$-$75  &$ -      $& 8&$ -5$& 1&dec &  436&M98  &$-V$ over profile\\
 	     &		  &$ -      $& 7&$ -7$& 1&dec &  600&C91  &$-V$ over profile\\
             &		  &$ -      $& 9&$ -8$& 1&dec &  660&Q95  &$-V$ over profile\\
             &		  &$ +-     $&11&$ -6$& 1&dec &  661&M98  &$+V$ for first 1/3 profile,$-V$ for rest\\
             & 		  &$ -      $&11&$ -9$& 2&dec &  950&v97  & similar at 800 MHz.\\   
             &		  &$ -      $& 9&$ -8$& 1&dec & 1440&Q95  &$-V$ over profile.\\
 0540$-$7125 &  	  &$ -/+    $&22&$  1$& 8&dec &  436&M98  &$-/+$ at center \\
 0543$+$2329 & 0540+23    &$ -      $&  &$   $&  &dec &  430&BR80 &$-V$ over profile (individual pulses)\\
             &		  &$ -      $&  &$   $&  &dec & 1400&R89  &$-V$ over profile.\\
             &		  &$ -      $&  &$   $&  &dec & 1720&Mr81 &as R89\\
             &		  &$ -      $&  &$   $&  &dec &10550&X95  &as R89\\
 0601$-$0527 & 0559$-$05  &$ -+     $&15&$ -3$& 3&inc & 1440&Q95  &bad s/n. $-V$, + V for 2 comp.  \\
             &		  &$ -+     $&  &$   $&  &inc & 1700&X91  &as Q95\\
 0614$+$2229 & 0611+22    &$ +      $&  &$   $&  &inc &  409&LM88 & $L\sim 100$\%.   \\
             &		  &$ +      $&  &$   $&  &inc &  430&RB81 &as LM88 at 409 MHz.   \\
 0629$+$2415 & 0626+24    &$ -/+     $&  &$   $&  &dec & 1400&R89  &$-/+$ for cone?  \\
 0630$-$2834 & 0628$-$28  &$ -      $&  &$   $&  &dec &  611&LM88 &$-V$ over profile. High $L$   \\
             &		  &$ -      $& 7&$ -7$& 2&dec &  649&Mc78 &as LM88, but bad s/n  \\
             &		  &$ -      $&11&$-10$& 1&dec &  950&v97  &as LM88 \\
             &		  &$ -      $&10&$ -9$& 2&dec & 1612&Ma80 &as LM88 \\
 0631$+$1036 &		  &$ ++     $&  &$   $&  &dec & 1418&Z96  &$+V$ for cone? $L\sim 100$\%.\\
             &		  &$ ++     $&  &$   $&  &dec & 1665&Z96  &as at 1418 MHz\\
 0659$+$1414 & 0656+14    &$ (+)-   $&  &$   $&  &dec & 1400&R89  &$+V$ not significant.\\
 0738$-$4042 & 0736$-$40  &$ -      $& 6&$ -5$& 2&xx  &  631&Mc78 &bad s/n and large dt. PA(f)?  \\
             &            &$ -      $& 7&$ -6$& 1&inc &  950&v97  &$-V$ stronger and over profile\\
             &            &$ -      $& 7&$ -6$& 1&inc & 1612&Ma80 &$-V$ for one comp.  \\
%\hline
\end{tabular}
\end{table*}\addtocounter{table}{-1}
\begin{table*}
\setlength{\tabcolsep}{2mm}
\caption{ Continued}
\begin{tabular}{lllrrrlrll}
\hline
%\hline
\multicolumn{1}{l}{PSR J}    &
\multicolumn{1}{l}{PSR B}    &
\multicolumn{1}{l}{$V$}& \relmodv & \relv & $\sigma$ &
\multicolumn{1}{l}{PA}     &
\multicolumn{1}{r}{Freq.}  &
\multicolumn{1}{l}{Ref.}   &
\multicolumn{1}{l}{Comments}\\
 &           &    &(\%) & (\%) & (\%) &    &(MHz)&  & \\
\hline
%
% NOTES for typesetter: Please always keep observations of one pulsar in one page, Never seperated !!!
%
 0742$-$2822 & 0740$-$28  &$ -      $& 2&$ -2$& 1&dec &  434&M98  &weak $-V$ over profile\\
             &            &$ -      $&10&$ -8$& 1&dec &  660&M98  &stronger $-V$ over profile\\
             &            &$ -      $& 6&$ -4$& 2&dec &  800&v97  &as 630 MHz by M78 but with bad dt \\
             &            &$ -      $& 8&$ -7$& 2&dec &  950&v97  &as 660 MHz of M98 but bad dt\\
             &            &$ -      $&  &$   $&  &dec & 1420&LM88 &$-V$ for cone?    \\
             &            &$ -      $& 8&$ -7$& 2&dec & 1612&Ma80 &as LM88 but with bad s/n.\\
 0754$+$3231 & 0751+32    &$ --     $&  &$   $&  &inc & 1400&R89  &$-V$ for two (conal?) comp \\
 0758$-$1528 & 0756$-$15  &$ --     $& 5&$ -5$& 2&inc & 1700&X91  &weak $-V$ for two comp \\
 0820$-$1350 & 0818$-$13  &$ -      $&19&$-19$& 4&xx  &  631&Mc78 &bad s/n\\
   	     &    	  &$ -      $&15&$-15$& 2&inc &  950&v97  & strong $-V$ for conal double?\\
 	     &   	  &$ -      $& 7&$ -5$& 1&inc & 1440&Q95  &$-V$ over profile. Bad dt\\
 0826$+$2637m& 0823+26m   &$ +/-    $&  &$   $&  &inc &  430&RB81 &$+/-$ for core \\
             & 	 	  &$ +/-    $&  &$   $&  &inc & 1400&R89  &$+/-$ for core  \\
             & 		  &$ -      $& 1&$ -1$& 0&inc & 1404&S84a & just weak $-V$ \\   
 0826$+$2637p& 0823+26p   &$ +      $&  &$   $&  &xx  &  430&RB81 &postcursor     \\
             & 		  &$ +      $&  &$   $&  &xx  & 1400&R89  &strong $+V$\\
 0828$-$3417 & 0826$-$34  &$ -+     $&  &$   $&  &d+i &  408&B85  &Wide profile, $-/+$ for last two comp\\
             & 	 	  &$ +-+    $&  &$   $&  &d+i &  610&B85  &1st comp is stronger \& shows $+V$  \\
 0835$-$4510 & 0833$-$45  &$ -      $& 6&$ -6$& 1&dec &  631&Mc78 &Vela. No $V$ at lower freq.\\ 
             & 		  &$ -      $& 9&$ -9$& 0&dec &  950&v97  &as Mc78 \\
             & 		  &$ -      $&14&$ 14$& 1&dec & 1612&Ma80 &as Mc78. Best data  \\
             & 		  &$ --     $&  &$   $&  &dec & 2295&KD83 &$-V$ for two components of Vela!   \\
 0837$+$0610 & 0834+06    &$ --     $&10&$-10$& 3&xx  &  636&Mc78 &$-V$ for 2 comps \\
             & 		  &$ +/--   $&12&$ -9$& 1&xx  &  800&S84b &$+/-$ for 1st comp, and $-V$ for 2nd comp\\
             & 		  &$ -      $& 9&$ -7$& 3&inc &  950&v97  & $-V$ only for 2nd comp \\
             & 		  &$ -+-    $& 5&$ -3$& 0&inc & 1404&S84a &$-V$ for comps, $+V$ for bridge.\\
             & 		  &$ -+-    $& 8&$ -7$& 2&inc & 1612&Ma80 &$-V$ for two comps\\
 0837$-$4135 & 0835$-$41  &$ +      $&27&$ 27$& 2&inc &  338&H77  &strong $+V$ for only comp\\
             & 		  &$ +      $&35&$ 35$& 3&inc &  405&H77  &strong $+V$ for only comp\\
             & 		  &$ +      $&20&$ 20$& 1&inc &  631&Mc78 &as H77. Better s/n. PA not clear. \\
             & 		  &$ +      $&30&$ 30$& 6&xx  &  800&v97  &PA dec? $V$ very strong\\ 
             & 		  &$ +      $&18&$ 17$& 2&dec &  950&v97  &PA jump? $V$ weaker\\  
             & 		  &$ +      $& 6&$  4$& 1&inc & 1612&Ma80 &very weak $+V$ only\\
 0840$-$5332 & 0839$-$53  &$ +      $&14&$  9$& 1&inc &  660&Q95  &$+V$ for first half profile.   \\
             &            &$ +      $&15&$ 10$& 4&inc & 1440&Q95  &$+V$ for central part, bad s/n\\
 0846$-$3533 & 0844$-$35  &$ -      $&19&$-16$& 3&dec &  950&v97  &PA dec, then flat? \\
             &            &$ -      $&12&$ -8$& 2&dec & 1440&Q95  &$-V$ with bad s/n.\\
 0907$-$5157 & 0905$-$51  &$ +(-)   $& 9&$  7$& 1&inc &  660&Q95  &$+V$ for central comp of high $L$\\
             &            &$ +-     $& 8&$  3$& 2&inc &  953&v97  &$+-V$ for two comp of partial cone?\\
 0908$-$4913 & 0906$-$49  &$ +      $&10&$  6$& 1&dec &  660&Q95  &weak $+V$ for main (stronger) pulse\\ 
 0922$+$0638 & 0919+06    &$ +      $& 9&$  7$& 2&inc &  950&v97  &$+V$ at strongest comp of profile\\
             &            &$ +      $& 9&$  9$& 1&inc & 1404&S84a &$+V$ at maximum \\
 0934$-$5249 & 0932$-$52  &$ +      $& 8&$  7$& 3&dec &  661&M98  &weak $+V$ over profile, bad s/n\\
  	     &            &$-       $&17&$-17$& 3&dec &  950&v97  & $-V$ over profile, bad dt. V(f)?\\
 0942$-$5552 & 0940$-$55  &$-       $&14&$ -2$& 4&inc & 1612&Ma80 & $-V$ over profile, bad s/n\\  
 0942$-$5657 & 0941$-$56  &$ +      $&17&$ 12$& 3&inc &  661&M98  &$+V$ over profile \\
  	     &            &$ +      $&23&$ 16$& 5&inc & 1411&Q95  &$+V$ for only comp. bad s/n, bad dt. \\
 0944$-$1354 & 0942$-$13  &$ -/+    $&22&$ 14$& 2&dec &  409&LM88 &core? Two comp at 1.42GHz (see S95)\\
 0946$+$0951 & 0943+10    &$ +      $&  &$   $&  &dec &  430&RB81 &$+V$ for the only comp \\
 0953$+$0755m& 0950+08m   &$ -      $&  &$   $&  &dec &  800&S84b &weak $-V$ for second half profile \\
             &            &$ -      $& 8&$ -6$& 0&dec &  950&v97  &$-V$ over profile\\
             &            &$ -      $&  &$   $&  &dec & 1400&R89  &weak V. V(f)?   \\
             &            &$ -      $& 6&$ -6$& 1&dec & 1404&S84a &$-V$ for whole profile. \\
             &            &$ -      $& 9&$ -9$& 1&xx  & 1612&Ma80 &worse s/n than S84a. \\
 0953$+$0755 & 0950+08i   &$ -+     $&  &$   $&  &dec & 1400&R89  &bad s/n \\
 0955$-$5304 & 0953$-$52  &$ +      $&13&$ 11$& 2&xx  &  661&M98  &$+V$ for strongest central component\\
 1001$-$5507 & 0959$-$54  &$ +/-    $& 4&$  2$& 1&inc &  654&M98  &weak $+/-$ for core, but PA not clear\\
  	     &            &$ +/-    $& 5&$  2$& 3&inc &  950&v97  &as 654MHz but bad s/n\\
 1034$-$3224 &            &$ +/-    $&  &$   $&  &xx  &  436&M98  &$+/-$ for strongest of more than 6 comp!\\
 1123$-$4844 &            &$ +      $&18&$ 13$& 5&dec?&  436&M98  &$+V$ strong between comps\\
%\hline
\end{tabular}
%\end{small}
\end{table*}\addtocounter{table}{-1}
\begin{table*}
%\begin{small}
\caption{ Continued}
\begin{tabular}{lllrrrlrll}
\hline
%\hline
\multicolumn{1}{l}{PSR J}    &
\multicolumn{1}{l}{PSR B}    &
\multicolumn{1}{l}{$V$}& \hspace{-4mm}\relmodv & \relv & $\sigma$ & 
\multicolumn{1}{l}{PA}     &
\multicolumn{1}{r}{Freq.}  &
\multicolumn{1}{l}{Ref.}   &
\multicolumn{1}{l}{Comments}\\
 &           &    &(\%) & (\%) & (\%) &    &(MHz)&  & \\
\hline
%
% NOTES for typesetter: Please always keep observations of one pulsar in one page, Never seperated !!!
%
 1136$+$1551 & 1133+16    &$ --     $&15&$-15$& 2&inc &  638&Mc78 &$-V$ for two comps and bridge! \\
             &            &$ --     $& 8&$ -5$& 4&inc &  950&v97  &bad s/n for $-V$\\
             &            &$ ---    $& 8&$ -8$& 1&inc & 1404&S84a &$-V$ is complicated for 1st comp.\\
             &            &$ --     $&10&$-10$& 1&inc & 1612&Ma80 &as Mc78.         \\
 1157$-$6224 & 1154$-$62  &$ +      $&20&$ 17$& 5&dec &  631&Mc78 &bad s/n for $V$  \\
             &            &$ +      $&15&$ 12$& 4&dec &  950&v97  &$+V$ for second part of profile\\
 1202$-$5820 & 1159$-$58  &$ +      $&10&$  9$& 2&inc &  436&M98  &$+V$ over profile \\
 1210$-$5559 &            &$ -      $& 8&$ -7$& 2&xx  &  436&M98  &$-V$ over profile, bad dt\\
 1224$-$6407 & 1221$-$63  &$ +      $&32&$ 15$& 7&inc &  631&Mc78 & bad s/n, strong $V$   \\
             &            &$ +      $&33&$ 26$& 5&inc &  950&v97  & Strong $+V$\\
             &            &$ +      $&14&$ 13$& 4&inc & 1560&W93  &s/n for $V$ just ok, but large dt. \\
             &            &$ +      $&20&$ 20$& 3&inc & 1612&Ma80 &$V$ is much weaker \\
 1239$+$2453 & 1237$+$25  &$ ++/--  $&  &$   $&  &dec &  408&LM88 &$+/-$ for core   \\
             &            &$ +++/-- $&  &$   $&  &dec &  410&M71  &$+/-$ for core, $+V$ for two conal comp\\
             &            &$ -/+/-  $&  &$   $&  &xx  &  430&B82  &normal mode, but ++0+ for abnormal mode\\
             &            &$ +++/-- $&13&$  4$& 1&dec & 1404&S84a &$V$ for cone \& core \\
             &            &$ ++/--  $&  &$   $&  &dec & 1700&B82  &same $V$ for both modes\\
             &            &$ +/-    $&13&$  0$& 4&xx  & 2700&Mr81 & \\
 1243$-$6423 & 1240$-$64  &$ +      $&19&$ 18$& 2&inc &  631&Mc78 &$+V$ for only comp \\
             &            &$ +      $& 9&$  8$& 3&inc &  800& v97 & weaker $+V$\\
             &            &$ -/+    $& 6&$  3$& 2&inc &  950& v97 & weaker $-/+V$, PA varies\\
             &            &$ -      $&13&$-13$& 2&inc & 1612&Ma80 & only $-V$, stronger\\
 1253$-$5820 &            &$ +      $&15&$ 10$& 3&inc &  436&M98  &$+V$ over profile \\
 1302$-$6350 & 1259$-$63  &$ +      $&13&$ 11$& 2&dec & 1520&MJ95 &$+V$ seen from one of two comp  \\
             &            &$ +      $&18&$ 12$& 3&dec & 4680&MJ95 &as at 1520 MHz\\
 1320$-$5359 & 1317$-$53  &$ -      $& 7&$ -5$& 7&dec &  600&C91  & weak $-V$ over profile\\ 
 1326$-$5859 & 1323$-$58  &$ -+     $&11&$  6$& 2&dec &  950&v97  &$-/+$ for core   \\
 1328$-$4357 & 1325$-$43  &$ -+     $&10&$  5$& 2&inc &  436&M98  &$-/+$ for first two of 3 comp   \\
 1328$-$4921 & 1325$-$49  &$ +/-    $&18&$  0$& 6&xx  &  436&M98  &$+/-$ near centre\\
 1338$-$6204 & 1334$-$61  &$ +      $&10&$  6$& 2&i+d & 1440&Q95  & weak $+V$ in central part\\   
 1341$-$6220 & 1338$-$62  &$ -      $&25&$-12$& 9&inc & 1411&Q95  &bad s/n for $-V$. $L\sim$90\%.\\
 1357$-$6228 & 1353$-$62  &$ -      $&16&$ 12$& 5&dec & 1612&Ma80 & bad s/n for $-V$ over profile\\
 1359$-$6038 & 1356$-$60  &$ +      $&22&$ 22$& 2&inc &  660&M98  &$+V$ over profile, $L\sim $ 100\%   \\
             &            &$ +-     $&15&$  4$& 4&inc & 1560&W93  &bad s/n for $-V$ and bad dt\\
 1453$-$6413 & 1449$-$64  &$ ++     $&11&$  9$& 1&inc &  400&H77  &$+V$ varies with freq.?\\
             &            &$ ++     $& 6&$  5$& 2&inc &  950&v97  & 3 components?\\
 1456$-$6843 & 1451$-$68  &$ -/+    $& 8&$  4$& 1&dec &  400&H77  &$-/+$ for core, same at 271 MHz (R83).  \\
             &            &$ --/++  $& 6&$  0$& 1&dec &  649&Mc78 &as at 400 MHz, $-/+$ at centre \\
             &            &$--/++   $& 6&$  3$& 0&dec &  950&v97  & two comp in core \\
             &            &$ --/++  $& 7&$ -1$& 1&dec & 1612&Ma80 & symmetric $V$?   \\
 1509$+$5531 & 1508+55    &$ +/-    $&  &$   $&  &dec &  610&LM88 &$+/-$ over triple profile  \\
             &            &$ ++/--  $& 8&$  0$& 2&dec & 1612&Mr81 & as 610 MHz \\
 1527$-$3931 & 1524$-$39  &$ +      $&18&$  8$& 6&dec &  436&M98  &$+V$ for conal double, bad s/n\\
             &            &$ +      $&20&$ 13$& 5&dec &  661&M98  &as at 436 MHz\\
 1527$-$5552 & 1523$-$55  &$ -/+    $&20&$ 13$& 2&dec &  661&M98  &$-/+$ for the core?  \\
 1534$-$5334 & 1530$-$53  &$ +/-    $&  &$   $&  &dec &  960&v97  & $+/-$for the first strong comp\\
             &            &$ +/-    $& 8&$ -4$& 2&dec & 1612&Ma80 & as v97, partial cone?\\
 1537$+$1155 & 1534+12    &$ --/+   $&  &$   $&  &dec &  430&A96  &$+/-$ for main comp, MSP \\
 1542$-$5304 &            &$ +      $&16&$ 10$& 5&inc &  661&M98  &$+V$ in 2nd half of profile\\
 1543$+$0929 & 1541+09    &$ +-+    $&  &$   $&  &d+i &  430&R83  &possible 7 comp, strong $-V$ at centre\\
             &            &$ +-+    $&13&$ -1$& 2&d+i & 1400&R89  &$V$ seen over the very wide profile.   \\
 1544$-$5308 & 1541$-$52  &$ --/++  $&11&$  1$& 2&xx  &  661&M98  &$-/+$ for central comp \\
 1557$-$4258 &            &$ -      $&13&$ -8$& 2&dec &  661&M98  &$-V$ for central comp with strong $L$\\
 1559$-$4438 & 1556$-$44  &$ -      $&12&$-12$& 2&dec &  631&Mc78 &$L\sim 60$\%\\
             &            &$ -+     $& 5&$  3$& 1&dec &  661&M98  &$V$ variable?   \\
             &            &$ -      $& 8&$ -7$& 4&dec &  800&v97  &weaker $V$\\
             &            &$ -      $&13&$-10$& 3&dec &  950&v97  &$-V$ stronger\\
             &            &$ -/+    $&11&$ -8$& 1&dec & 1490&M98  &resolved profile with interesting $V$  \\
             &            &$ -      $& 8&$ -4$& 3&dec & 1560&W93  &weak $-V$.  \\
             &            &$ -      $&16&$-15$& 2&dec & 1612&Ma80 &$-V$ at centre\\
%\hline
\end{tabular}
%\end{small}
\end{table*}\addtocounter{table}{-1}
\begin{table*}
\begin{small}
\caption{ Continued}
\begin{tabular}{lllrrrlrll}
\hline
%\hline
\multicolumn{1}{l}{PSR J}    &
\multicolumn{1}{l}{PSR B}    &
\multicolumn{1}{l}{$V$}& \relmodv & \relv & $\sigma$ &
\multicolumn{1}{l}{PA}     &
\multicolumn{1}{r}{Freq.}  &
\multicolumn{1}{l}{Ref.}   &
\multicolumn{1}{l}{Comments}\\
 &           &    &(\%) & (\%) & (\%) &    &(MHz)&  & \\
\hline
%
% NOTES for typesetter: Please always keep observations of one pulsar in one page, Never seperated !!!
%
 1600$-$5044 &1557$-$50   &$ +      $&21&$ 20$& 5&inc &  950&v97  & bad s/n, scattering\\
             &            &$ +      $&16&$ 14$& 2&inc & 1560&W93  &bad dt, $+V$ in part of profile\\
             &            &$ +      $&23&$ 21$& 4&inc & 1612&Ma80 &as W93.\\
 1604$-$4909 &           &$ +/-    $& 6&$  3$& 1&inc &  661&M98  &$+/-$ for core  \\
 1607$-$0032 &1604$-$00  &$ -/++   $&  &$   $&  &dec &  430&R88  &$-/+$ for one comp, $+V$ for another comp\\
             &           &$  -     $& 9&$ -7$& 2&xx  &  631&Mc78 &$-V$ over most of profile\\
             &           &$ +-+    $&  &$   $&  &xx  & 1400&R89  &$+V$ doubtful, $-V$ OK \\
 1614$-$5047 &1610$-$50  &$  +     $&26&$ 10$&12&dec & 1411&Q95  &bad dt \\
 1633$-$5015 &1629$-$50  &$  -     $&18&$-13$& 3&inc & 1411&Q95  &strong $-V$ over profile  \\
 1635$+$2418 &1633+24    &$  -(+)  $&  &$   $&  &dec & 1400&R89  &$-V$ for central comp\\
 1644$-$4559 &1641$-$45  &$ -      $& 3&$ -2$& 0&inc &  950&v97  & weak $-V$ over scattered profile\\   
             &           &$ (-)+-  $& 4&$ -1$& 1&inc & 1612&Ma80 &$+V$ for peak. \\
 1646$-$6831 &1641$-$68  &$ +(-)   $&12&$ 5 $& 1&i+d &  660&Q95  &$+V$ at centre  \\
             &           &$ +(-)   $&14&$ 5 $& 3&i+d &  953&v97  &bad s/n for $-V$, PA = inc+dec?\\
 1645$-$0317 &1642$-$03  &$ +/-    $&  &$   $&  &xx  &  408&LM88 &complicated PA  \\
             &           &$ -      $&  &$   $&  &xx  &  410&M71  &$-V$ over profile\\ 
             &           &$ -      $& 8&$ -7$& 1&inc &  631&Mc78 &$-V$ for the only comp, V(f)?   \\
             &           &$ -      $& 3&$ -2$& 1&dec &  950&v97  &$-V$ for the only comp, PA(f)?\\
 1651$-$4246 &1648$-$42  &$ -      $&19&$-13$& 4&dec &  950&v97  &$-V$ for 2nd part of profile, bad s/n.\\  
             &           &$ -      $&11&$ -3$& 4&dec & 1560&W93  &as at 950 MHz, better s/n \\
 1700$-$3312 &           &$-       $&27&$-24$& 8&inc &  434&M98  &$-V$ over profile, but with bad s/n\\
 1703$-$3241 &1700$-$32  &$ -/+    $& 7&$ -1$& 2&inc &  950&v97  &$-/+$ for core\\
             &           &$ -/+    $&10&$ -2$& 3&inc & 1612&Ma80 &$-/+$ for core seen at 409MHz (LM88).\\
 1705$-$1906 &1702$-$19m &$ -      $&60&$-60$& 3&dec &  408&B88  &large $-V$ over whole profile \\
 1705$-$1906 &1702$-$19i &$ -      $&60&$-60$& 3&dec &  408&B88  &also large $-V$ \\
 1709$-$4428 &1706$-$44  &$ -      $&17&$-16$& 3&inc & 1440&Q95  &$-V$ over only comp, $L$ up to 90\%.   \\
 1713+0747   &           &$ -/+    $&  &$   $&  &i+d & 1400&XK96 &$-/+$ for core? PA inc+dec?\\
 1721$-$3532 &1718$-$35  &$ -      $&17&$-14$& 4&dec & 1411&Q95  &$-V$ appears at strongest part of profile\\
 1722$-$3712 &1719$-$37  &$ +      $& 9&$  8$& 2&inc &  661&M98  &weak $+V$ over profile, high $L$\\
 1731$-$4744 &1727$-$47  &$ +      $& 9&$  7$& 3&xx  &  400&H77  & bad s/n and PA not clear\\ 
             &           &$ +      $& 5&$  4$& 1&dec &  800&v97  & weak $+V$ for 1st comp. Same at 950 MHz\\
 1740$-$3015 &1737$-$30  &$ -      $&20&$-19$& 2&dec &  660&Q95  &$-V$ over profile, bad dt \& bad s/n   \\
             &           &$ -      $&22&$-22$& 3&dec & 1560&W93  &$-V$ strong at 2nd half of only comp.\\
 1740$+$1311 &1737+13    &$ ++/-   $&  &$   $&  &dec &  430&R88  &$+/-$ for core \& $+V$ for cone \\
             &           &$ +++/-  $&  &$   $&  &dec & 1400&R89  &as R88 \\
 1741$-$3927 &1737$-$39  &$ +      $& 7&$  4$& 1&xx  &  661&M98  &weak $+V$ over profile   \\
 	     &           &$ +      $&10&$  7$& 3&inc &  954&v97  &$+V$ at centre, bad dt\\
 1745$-$3040 &1742$-$30  &$ -      $&10&$ -9$& 3&xx  &  950&v97  & PA and V similar to X91 \\
             &           &$ ---    $&15&$-15$& 2&xx  & 1700&X91  &$-V$ for 3 (conal?) comp   \\
 1751$-$4657 &1747$-$46  &$ -/++   $&12&$ 8 $& 1&dec &  436&M98  &$-/+$ for 1st comp, 2nd comp $+V$ \\
 	     &           &$   +    $& 8&$ 6 $& 1&dec &  631&Mc78 &$+V$ at center\\
 	     &           &$ -/++   $&  &$   $&  &dec &  950&v97  & as M98 at 1522MHz.\\
             &           &$ -/++   $& 9&$ 6 $& 2&dec & 1522&M98  &same as 436 MHz  \\
 1752$-$2806 &1749$-$28  &$ -      $& 7&$ -7$& 1&xx  &  631&Mc78 &bad dt, V(f)?  \\
             &           &$ -      $& 3&$ -2$& 0&dec &  950&v97  & V(f), PA(f)\\
             &           &$ -(/+)  $& 5&$  1$& 1&xx  & 1612&Ma80 &weak $+V$, bad dt.  \\
             &           &$ -/+    $& 5&$  0$& 3&xx  & 1720&Mr81 &$-/+$ for core? \\
             &           &$ -/+    $&13&$ 11$& 3&dec & 2650&Mr81 &$-/+$ for core? 3 comp?\\
 1801$-$0357 &           &$ +/-    $&20&$ -6$& 4&dec &  661&M98  &$+/-$ for core, but PA not clear\\
 1803$-$2137 &1800$-$21  &$ ++     $&21&$ 20$& 3&dec & 1560&W93  &interesting profile with high $L$\\
 1807$-$0847 &1804$-$08  &$ +--    $&10&$  0$& 2&xx  & 1700&X91  & very weak V for central and last comp\\
 1817$-$3618 &1813$-$36  &$ -      $&13&$-10$& 1&xx  &  660&Q95  &strong $-V$ over part profile, bad dt\\
 1820$-$0427 &1818$-$04  &$ -      $&  &$   $&  &xx  &  410&M71  &$-V$ over part profile  \\
             &           &$ -      $&18&$-12$& 4&inc &  631&Mc78 &poor s/n  \\
             &           &$ -      $&13&$ -6$& 3&inc &  950&v95  &better s/n\\
 1823+0550   & 1821+05   &$ ++/--  $&11&$  2$& 2&dec & 1400&R89  &$+/-$ for core, PA not clear\\
 1829$-$1751 & 1826$-$17 &$ -      $&19&$-17$& 2&xx  &  436&M98  &strong $-V$, scatterred profile. Flat PA\\
             &           &$ -      $&15&$-11$& 2&d+i &  660&M98  &strong $-V$ over profile, PA= dec+inc!\\
             &           &$ -      $&17&$-11$& 5&xx  &  950&v97  & PA inc+dec?\\
 1836$-$1008 & 1834$-$10 &$ -+     $&10&$ -3$& 3&dec & 1720&X91  &weak $-/+$ at pulse center. poor s/n \\
 1841+0912   & 1839+09   &$ +-+    $& 5&$ -3$& 1&dec & 1400&R89  &$+-+$ for 3 comp, $+V$ very weak.\\
 1848$-$1952 & 1845$-$19 &$ -      $&14&$-13$& 5&xx  &  950&v97  & $-V$ for 1st resolved component\\
%\hline
\end{tabular}
\end{small}
\end{table*}\addtocounter{table}{-1}
\begin{table*}
\begin{small}
\caption{ Continued}
\begin{tabular}{lllrrrlrll}
\hline
%\hline
\multicolumn{1}{l}{PSR J}    &
\multicolumn{1}{l}{PSR B}    &
\multicolumn{1}{l}{$V$}& \relmodv & \relv & $\sigma$ &
\multicolumn{1}{l}{PA}     &
\multicolumn{1}{l}{Freq.}  &
\multicolumn{1}{l}{Ref.}   &
\multicolumn{1}{l}{Comments}\\
 &           &    &(\%) & (\%) & (\%) &    &(MHz)&  & \\
\hline
%
% NOTES for typesetter: Please always keep observations of one pulsar in one page, Never seperated !!!
%
 1852$-$2610 &           &$   +    $& 9&$  2$& 3&dec &  436&M98  &strong $L$, weak $V$ \\
 1857$+$0943 &1855+09m   &$  +-    $&  &$   $&  &dec & 1400&S86  &+$-V$ for 2 comp. \\
 1900$-$2600 &1857$-$26  &$ +/-    $&  &$   $&  &xx  &  170&R83  &$+/-$ for core comp  \\
             &           &$ +/--   $&  &$   $&  &xx  &  268&R83  &$+/-$ for core comp\\
             &           &$ ++/--  $&17&$ -3$& 4&dec &  631&Mc78 &$+/-$ for central part \\
             &           &$ +/-    $&12&$  2$& 2&dec &  661&M98  &$+/-$ for central part  \\
             &           &$ +/-    $&12&$ -1$& 1&dec &  950&v97  &$+/-$ for central part  \\
             &           &$ +/-    $&16&$ -2$& 1&dec & 1490&M98  &$+/-$ for central part  \\
             &           &$ ++/--  $&20&$ -1$& 3&dec & 1612&Ma80 &$+/-$ over most of profile  \\
             &           &$  +/-   $&22&$ 0 $& 4&dec & 2650&Mr81 & \\
 1901$+$0331 &1859+03    &$ +/-    $&12&$ 3 $& 2&dec & 1400&R89  &$+/-$ for core\\
 1903$+$0135 &1900$+$01  &$ +      $&13&$  8$& 4&inc &  950&v97  & $+V$ for central comp only\\
 1907$+$4002 &1905+39    &$ --     $&10&$-10$& 3&dec & 1700&X91  &$-V$ for 2 comp, s/n just ok.  \\
 1909$+$0254 &1907+02    &$ -/+    $&  &$   $&  &dec &  430&RB81 &triple at 1410 MHz (S95) \\
 1910$+$0358 &1907+03    &$ (-+)-  $&16&$-13$& 2&dec & 1400&R89  &strong $-V$ from 2nd half profile.\\
 1909$+$1102 &1907+10    &$ -      $&12&$-12$& 2&inc & 1400&R89  &$-V$ for core only? \\
 1910$+$1231 &1908+12    &$ +      $&  &$   $&  &xx  &  430&RB81 &s/n for $+V$ just ok\\
 1912+2104   & 1910+20   &$ -      $&  &$   $&  &inc & 1400&R89  &s/n for $-V$ just ok\\
 1913$-$0440 &1911$-$04  &$ +-+    $&15&$ -1$& 4&xx  &  400&H77  &bad s/n for $V$ \\
             &           &$ +-     $& 8&$ -5$& 2&inc &  950&v97  & $+-$ for two unresolved comp? \\
 1913+1400   & 1911+13   &$ +-(+)  $&  &$   $&  &xx  & 1400&R89  &bad s/n for $+V$. PA=dec?\\
 1915+1009   & 1913+10   &$ +      $&40&$ 40$& 2&d+i & 1400&R89  &very strong $+V$ over profile\\ 
 1915+1606   & 1913+16   &$ -/++-  $&  &$   $&  &inc &  430&C90  &$-/+$ for 1st comp\\
 1915+1647   & 1913+167  &$ +      $&  &$   $&  &dec &  430&RB81 & $+V$ at centre\\
             &           &$ +/-    $&  &$   $&  &xx  & 1400&R89  &two core comps \\
 1916+0951   & 1914+09   &$ -      $&  &$   $&  &dec &  430&RB81 &$-V$ for 1st comp\\
             &           &$+       $&  &$   $&  &dec & 1400&R89  &weak $+V$ over profile\\  
 1916+1312   & 1914+13   &$ -      $&37&$-37$& 2&inc & 1400&R89  &very strong $-V$ over profile\\
 1917+1353   & 1915+13   &$ -      $&  &$   $&  &dec &  430&RB81 &$-V$ for only comp   \\
             &           &$ -      $&  &$   $&  &dec & 1400&R89  &$-V$ over profile  \\
 1918+1444   & 1916+14   &$ +      $&  &$   $&  &xx  &  430&RB81 &bad s/n\\
             &           &$ +      $&  &$   $&  &dec & 1400&R89  &$+V$ over profile  \\
 1919+0021   & 1917+00   &$ +      $&  &$   $&  &xx  &  418&RB81 &$+V$ for core \\
             &           &$ ++-    $&  &$   $&  &inc & 1400&R89  &triple profile, $+/-$?\\
 1921+1419   & 1919+14   &$ +-     $&  &$   $&  &dec &  430&RB81 &$+V$ weak, $-V$ strong.   \\
             &           &$ +-     $&  &$   $&  &dec & 1400&R89  &$-V$ not so strong\\
 1922+2110   & 1920+21   &$ +      $&20&$ 3 $& 2&dec &  430&RB81 &$+V$ for core? bad dt.   \\
 1926+1648   & 1924+16   &$ -/+    $&  &$   $&  &inc &  430&RB81 & $-/+$? high $L$\\
             &           &$ +      $&  &$   $&  &inc & 1400&R89  & weak $+V$ over profile\\   
 1932+1059m  & 1929+10m  &$ --     $&  &$   $&  &dec &  430&BR80 &$-V$ for all comps, $L\sim 100$\%\\
             &           &$ ++--   $& 2&$  0$& 0&dec &  800&S84b &$V$ very weak, $+/-V$ for central comp.\\
             &           &$ --     $&  &$   $&  &dec & 1400&R89  &weaker $-V$ for 2 or 3 comp.   \\
             &           &$ --     $& 1&$ -1$& 0&dec & 1404&S84a &weaker $-V$ for 3 comp.\\
             &           &$ --     $&  &$   $&  &dec & 1665&P90  &$-V$ for all at 1665 and 430 MHz.   \\
             &           &$ ++     $&  &$   $&  &dec & 1700&X91  &very weak $+V$ for all   \\
 1932+1059m  & 1929+10i  &$ -      $&  &$   $&  &inc &  430&RB81 &$-V$ for main comp   \\
             &           &$ -      $&  &$   $&  &xx  & 1400&R89  &$-V$ for core?\\
             &           &$ -      $&  &$   $&  &inc & 1665&P90  &$-V$ for all at 430 and 1665 MHz.\\
 1932$-$3655 &           &$ +/-    $&22&$ 9 $& 8&inc &  658&M98  &significant $+/-V$ for the peak comp\\
 1935+1616   & 1933+16   &$ -/+    $&  &$   $&  &xx  &  430&R83  &$-/+$ for core  \\
             &           &$ -/+    $&15&$ -1$& 1&xx  & 1440&R89  &$-/+$ for two core comps\\
             &           &$ -/+    $&14&$ 0 $& 3&xx  & 1720&Mr81 &good s/n\\
             &           &$ -/+    $&15&$ 0 $& 3&xx  & 2700&Mr81 &$-/+$ for core  \\
 1939+2134m  &1937+21m   &$ -      $&  &$   $&  &xx  & 1418&TS90 & across whole pulse\\
 1939+2134i  &1937+21i   &$ -      $&  &$   $&  &xx  & 1418&TS90 & \\
 1941$-$2602 &1937$-$26  &$ -      $&13&$ -7$& 5&dec & 1560&W93  &weak $-V$, s/n just ok\\
 1946+1805   & 1944+17   &$ -(+)-- $& 5&$ 4 $& 1&inc & 1404&S84a &$V$ over profile   \\
 1946+2244   & 1944+22   &$ -      $&  &$   $&  &xx  &  430&RB81 &s/n just ok \\
 1946$-$2913 &1943$-$29  &$ +      $&15&$ 9 $& 6&inc &  434&M98  & $+V$ just for central strong comp\\ 
 1948+3540   & 1946+35   &$ +      $&  &$   $&  &xx  &  430&RB81 &$+V$ for scattered profile \\
             &           &$ ++     $&  &$   $&  &inc & 1400&R89  &$+V$ for 2 or 3 comp \\
%\hline
\end{tabular}
\end{small}
\end{table*}\addtocounter{table}{-1}

\begin{table*}
\begin{small}
\caption{ Continued}
\begin{tabular}{lllrrrlrll}
\hline
%\hline
\multicolumn{1}{l}{PSR J}    &
\multicolumn{1}{l}{PSR B}    &
\multicolumn{1}{l}{$V$}& \relmodv & \relv & $\sigma$ &
\multicolumn{1}{l}{PA}     &
\multicolumn{1}{l}{Freq.}  &
\multicolumn{1}{l}{Ref.}   &
\multicolumn{1}{l}{Comments}\\
 &           &    &(\%) & (\%) & (\%) &    &(MHz)&  & \\
\hline
%
% NOTES for typesetter: Please always keep observations of one pulsar in one page, Never seperated !!!
%
 1954+2923   & 1952+29   &$ --+-   $&  &$   $&  &inc &  430&RB81 &strong $-V$ \\
             &           &$ ---+-  $&21&$-19$& 2&de? & 1400&R89  &$-V$ over most of profile  \\
 1959+2048m & 1957+20m   &$ (+)/-  $&  &$   $&  &xx  &  430&TS90 &MSP. very weak $L$\\
            &            &$ -/+    $&  &$   $&  &xx  &  430&F90  &$V$ appears to be reversed to TS90.\\
 1959+2048i &1957+20i    &$ +-     $&  &$   $&  &xx  &  430&TS90 &as main pulse\\
            &            &$ -+     $&  &$   $&  &xx  &  430&F90  &as main pulse\\
 2004+3137  & 2002+31    &$ +/-    $&  &$   $&  &xx  &  430&RB81 &$+/-$ for core,  bad dt.\\
            &            &$ +/-    $&4 &$ 0 $& 1&inc & 1400&R89  &PA is not so clear.\\
 2006$-$0807&2003$-$08   &$ -+/--  $&14&$-11$& 2&inc & 1700&X91  &$+/-$ core, $-V$ for cone \\
 2018+2839  & 2016+28    &$ -      $& 6&$ -6$& 1&inc & 1404&S84a &$-V$ for central part   \\
 2022+2854  & 2020+28    &$ +/---  $&  &$   $&  &inc &  430&C78  &But also show $--$ for two cones!   \\
            &            &$ +/---  $& 8&$ -7$& 0&inc &  800&S84b &$+/-$ for 1st comp \\
            &            &$ +/---  $& 7&$ -7$& 0&inc & 1404&S84a &as at 800 MHz, but $+/-$ weak\\
 2022+5154  & 2021+51    &$ -      $& 4&$ -4$& 1&inc &10550&X95  &weak $-V$    \\
 2021+2145  & 2025+21    &$ +      $&  &$   $&  &dec &  430&RB81 &s/n just ok   \\
 2030+2228  & 2028+22    &$ ++     $&  &$   $&  &dec & 1400&R89  &weak $+V$ for 2 comps\\
 2046$-$0421&2043$-$04   &$ -      $&  &$   $&  &xx  &  950&v97  & $-V$ for strongest comp\\
 2046+1540  & 2044+15    &$ +++    $&  &$   $&  &i+d & 1400&R89  & very weak $+V$ for 2 comp\\
 2048$-$1616& 2045$-$16  &$ +-+    $& 7&$  3$& 1&dec &  950&v97  & $+V$ for outer conal comps\\
            &            &$ ++/-+  $&  &$   $&  &dec & 1420&LM88 & weak $+/-$ for core, $+V$ for conal comps\\
 2053$-$7200&2048$-$72   &$ +-     $&13&$ -4$& 1&inc &  660&Q95  & $+-V$ for 2 comp with bad s/n. \\
  	    &            &$ +/-    $& 8&$ -2$& 2&inc &  661&M98  &confirm $+-V$ seen by Q95\\
            &            &$ +-     $&  &$   $&  &inc &  950&v97  &$+-$ V for two comps\\
            &            &$ -+     $&10&$ -5$& 2&inc & 1440&Q95  &bad s/n for $V$, $V$(f)?\\
 2055+3630  & 2053+36    &$ -      $&  &$   $&  &xx  & 1400&R89  &bad s/n \\
 2113+2754  & 2110+27    &$ --     $&  &$   $&  &dec & 1400&R89  &weak $V$\\
 2113+4644  & 2111+46    &$ +/-    $&  &$   $&  &dec &  610&LM88 &$+/-$ for core \\
            &            &$ +/-    $&10&$ 0 $& 3&dec & 1720&Mr81 &also 2650 MHz.  \\
 2116+1414  & 2113+14    &$ +/-    $&  &$   $&  &inc & 1400&R89  &$+/-$ at centre but not peak\\
 2144$-$3933&            &$ -/+    $&17&$ 3 $& 2&inc &  661&M98  &$-/+$ for core  \\
 2145$-$0750&            &$ -+     $&  &$   $&  &in? & 1400&XK96 & PA variation not clear.\\
 2155$-$3118& 2152$-$31  &$ +      $&25&$ 22$& 7&de? &  950&v97  & Strong $+V$, weak $L$\\
 2219+4754  & 2217+47    &$ +      $&  &$   $&  &inc & 1612&Mr81 &but at 2650 MHz, weak $-V$?  \\
 2225+6535  & 2224+65    &$ --     $&  &$   $&  &dec &  408&LM88 &$-V$ for 2 comps, bad s/n, PA=dec+flat\\
 2305+3100  & 2303+30    &$ ++     $&  &$   $&  &inc & 1400&R89  &weak $+V$ over comp, good s/n\\
 2313+4253  & 2310+42    &$ -+-    $&13&$-11$& 1&inc & 1700&X91  &$V$ over profile\\
 2317+2149  & 2315+21    &$ -      $&  &$   $&  &inc &  409&LM88 &core? bad dt. bad s/n \\
            &            &$ ---    $&  &$   $&  &xx  & 1400&R89  &$-V$ varies for 2 (or 3) comp  \\
 2324$-$6054&2321$-$61   &$ +      $&13&$ 11$& 1&dec &  660&Q95  &$+V$ for 1 of 2 (conal) comp  \\
            &            &$ +(-)   $&  &$   $&  &dec &  950&v97  & bad s/n for $V$\\ 
 2330$-$2005& 2327$-$20  &$ --     $&22&$-22$& 1&inc &  648&Mc78 &strong $-V$ over whole profile\\
            &            &$ --     $&12&$-11$& 1&inc &  950&v97  &weaker $-V$, good s/n\\
 2346$-$0609&            &$ +      $&11&$  9$& 2&dec &  436&M98  &conal double, $+V$ for second comp \\
 2354+6155  & 2351+61    &$ ++     $& 9&$  9$& 3&dec & 1700&X91  &$+V$ for two comp\\
\hline
\hline
\multicolumn{10}{l}{\parbox{177mm}{
References: 
Manchester 1971 (M71);
Hamilton et al. 1977 (H77);
Cordes, Rankin \& Backer 1978 (C78);
McCulloch et al. 1978 (Mc78);
Backer \& Rankin 1980 (BR80);
Manchester, Hamilton \& McCulloch 1980 (Ma80);
Morris et al. 1981 (Mr81);
Rankin \& Benson 1981 (RB81);
Barter et al. 1982 (B82);
Krishnamohan \& Downs 1983 (KD83);
Rankin 1983 (R83);
Stinebring et al. 1984a (S84a); Stinebring et al. 1984b (S84b);
Biggs et al. 1985 (B85);
Segelstein et al. 1986 (S86);
Biggs et al. 1988 (B88);
Lyne \& Manchester 1988 (LM88);
Rankin, Wolszczan \& Stinebring 1988 (R88);
Rankin, Stinebring \& Weisberg 1989 (R89);
Cordes, Wasserman \& Blaskiewicz 1990 (C90);
Fruchter et al. 1990 (F90);
Phillips 1990 (P90);
Thorsett \& Stinebring 1990 (TS90);
Costa et al. 1991 (C91);
Xilouris et al. 1991 (X91);
Wu et al. 1993 (W93);
Gould 1994 (G94);
Manchester \& Johnston 1995 (MJ95);
Qiao et al. 1995 (Q95);
Seiradakis et al. 1995 (S95);
Xilouris et al. 1995 (X95);
Arzoumanian et al. 1996 (A96);
Xilouris \& Kramer 1996 (XK96);
Zepka, Cordes \& Wasserman 1996 (Z96);
Navarro et al. 1997 (N97);
van Ommen et al. 1997 (v97);
Manchester, Han \& Qiao 1998 (M98).
}} \\
\end{tabular}
\end{small}
\end{table*}